\documentclass[prd,aps,twocolumn,a4paper,floatfix,amsmath,amssymb]{revtex4-1}
\usepackage{graphicx,color}

\begin{document}

\title{The Einstein-Vlasov system in spherical symmetry: reduction of
  the equations of motion and classification of single-shell static
  solutions, in the limit of massless particles}

\author{Carsten Gundlach} 

\affiliation{Mathematical Sciences,
University of Southampton, Southampton SO17 1BJ, United Kingdom}

\date{27 October 2016, revised 7 December 2016}

\begin{abstract}

We express the Einstein-Vlasov system in spherical symmetry in terms
of a dimensionless momentum variable $z$ (radial over angular
momentum). This regularises the limit of massless particles, and in
that limit allows us to obtain a reduced system in independent
variables $(t,r,z)$ only. Similarly, in this limit the Vlasov density
function $f$ for static solutions depends on a single variable $Q$
(energy over angular momentum). This reduction allows us to show that
any given static metric which has vanishing Ricci scalar, is vacuum at
the centre and for $r>3M$ and obeys certain energy conditions uniquely
determines a consistent $f=\bar k(Q)$ (in closed form). Vice versa, any
$\bar k(Q)$ within a certain class uniquely determines a static metric
(as the solution of a system of two first-order quasilinear
ODEs). Hence the space of static spherically symmetric solutions of
Einstein-Vlasov is locally a space of functions of one variable. For a
simple 2-parameter family of functions $\bar k(Q)$, we construct the
corresponding static spherically symmetric solutions, finding that
their compactness is in the interval $0.7\lesssim \rm max_r(2M/r)\le
8/9$. This class of static solutions includes one that agrees with the
approximately universal type-I critical solution recently found by
Akbarian and Choptuik (AC) in numerical time evolutions. We speculate on
what singles it out as the critical solution found by fine-tuning
generic data to the collapse threshold, given that AC also found
that {\em all} static solutions are one-parameter unstable and sit on
the threshold of collapse.

\end{abstract}

\maketitle

\tableofcontents

\section{Introduction}

This work is an attempt to better understand some interesting recent
numerical results by Akbarian and Choptuik \cite{AkbarianChoptuik}
(from now on, AC) on type-I critical collapse in the spherically
Einstein-Vlasov system with massless particles. (For a general review
of critical collapse, see \cite{GundlachLRR}. Critical collapse in the
same system with massive particles was considered in
\cite{AndreassonRein2006}.)  We begin with a brief summary of what we
consider their two key results.

(I) Evolving several families of initial data for the massless
spherically symmetric Einstein-Vlasov system, AC found that they could
fine-tune any one parameter of these initial data sets to the collapse
threshold. They found type-I critical collapse with a static critical
solution that has an approximately universal metric after rescaling to
$M=1$, but not a universal matter distribution $f$. This critical
metric has compactness $\Gamma\simeq0.80\pm 0.01$ and central
redshift $Z_c\simeq2.45\pm 0.05$. The lifetime scaling exponent is
also approximately universal with value $\sigma\simeq 1.4\pm 0.1$. It
is not clear if the small variations of the critical spacetime and
critical exponents are numerical error or a genuine variation.

(II) In a separate series of numerical experiments, AC constructed static
solutions by specifying the matter distribution and solving for the
metric. The resulting metrics are clearly not universal, with
compactness and central redshift in the ranges $0.80\lesssim
\Gamma\lesssim 0.89$ and $2.0\lesssim Z_c\lesssim 2.4$. AC then evolved
these static solutions plus a small perturbation from one of three
families. They found that {\em all} their static solutions are at the
threshold of collapse, in the sense that for one sign of the
perturbation they collapse and for the other sign they disperse. They
also found lifetime scaling with $\sigma\simeq1.43\pm 0.07$, consistent
with the existence of a single unstable mode with a universal value of
its Lyapunov exponent equal to the one seen in fine-tuned generic
initial data.

In this paper we clarify the nature of the space of static spherically
symmetric massless solutions, leaving their perturbations to future
work.

In Sec.~\ref{section:newform} we cast the spherical Einstein-Vlasov
equation in a new form, based on an idea in \cite{critvlasov}. (In the
meantime, a similar formulation has been given in
\cite{RendallVelazquez2}.) The key idea is to introduce independent
variables $(t,r,z,F)$ such that the particle mass $m$ and particle
angular momentum squared $F$ appear only in the combination
$m^2/F$. The purpose of this section is to establish notation and to
demonstrate that the new formalism is as natural as the standard one
for $m>0$. In Sec.~\ref{section:massless}, we take the limit $m=0$, in
which $F$ can be integrated out. We then restrict this reduced system
to static solutions and show that given one of the following three
functions -- a static density function $f=\bar k(Q)$, or one of the
two metric coefficients $a_0(r)$, $\alpha_0(r)$ of a generic static
spherically symmetric metric -- one can uniquely obtain the other two.

In Sec.~\ref{section:examples} we numerically construct examples of
static massless solutions for a simple 2-parameter family of functions
$\bar k(Q)$. In Sec.~\ref{section:solutionspace} we formalise our
theoretical and numerical insights into three conjectures about the
space of static solutions. One particular static solution we have
constructed agrees well with the (approximately) universal critical
solution found by AC in their time evolutions, to within their
numerical accuracy. In Sec.~\ref{section:criticalsolution} we
speculate on what singles out this critical solution --- actually a
class of solutions with the same leading order behaviour --- from all
other static solutions.

In Sec.~\ref{section:conclusions} we summarise, and discuss the
relation of our results to those of AC.

Appendixes~\ref{appendix:scaleinvariance} and
\ref{appendix:u1expansions} contain technical details that would
disrupt the main argument, and Appendixes \ref{appendix:ACansatz} and
\ref{appendix:AFTansatz} express other ans\"atze for static solutions
from the literature in our notation. Throughout this paper, $x:=y$
means that $x$ is being defined, and the suffix $0$ denotes static
solutions. 

\section{The Einstein-Vlasov system in spherical symmetry}
\label{section:newform}

\subsection{Field equations}

The Einstein-Vlasov system describes the evolution of a statistical
ensemble of non-interacting particles coupled to gravity through their
average stress-energy. We restrict ourselves to a single species of
particles with mass $m\ge 0$. 

We describe the state of a many-body system with a distribution
function over the phase space of the system. As the particles are
equivalent and uncorrelated, we can use the one-particle distribution
function $f(x^\mu, p_\nu)$, where $x^\mu$ are coordinates in spacetime
and $p_{\mu}$ are coordinates in the cotangent space at that
point. Because of the constraint $p^\mu p_\mu=-m^2$, we can consider
the distribution $f$ as a function of $(x^\mu, p^i)$.

Each particle follows a geodesic. This defines a congruence of curves
on phase space, tangent to the Liouville operator (or ``geodesic
spray'')
\begin{eqnarray}
\frac{d}{d\sigma}
  &:=& \frac{dx^\mu}{d\sigma} \frac{\partial}{\partial x^\mu}
  + \frac{dp^i}{d\sigma} \frac{\partial}{\partial p^i} \\
  &=& p^\mu \frac{\partial}{\partial x^\mu} -
  \Gamma^i_{\nu\lambda} p^\nu p^\lambda \frac{\partial}{\partial p^i}.
\end{eqnarray}
Here $\sigma$ is an affine parameter related to proper time
$s$ through $ds=m d\sigma$ for massive particles. This allows us to
treat massless particles as the limit $m\to 0$ of massive ones.  As
the particles do not interact directly, the evolution of $f$ is
governed by the Vlasov equation
\begin{equation}
{\cal L}f=0 .
\label{Vlasov}
\end{equation}

The Einstein equations are $G_{\mu\nu}=8\pi T_{\mu\nu}$, with the
stress-energy tensor of collisionless matter given by
\begin{equation}
\label{Tmunuf}
T_{\mu\nu}=\int p_\mu p_\nu f\,  dV_p .
\end{equation}
The invariant volume element on the mass shell in 4-momentum space can
be written as
\begin{equation}
\label{dVp}
dV_p= 2\sqrt{-g}\,\delta(p_\mu p^\mu+m^2)\,d^4p^\mu 
={\sqrt{-g}\,d^3p^i \over p_t}.
\end{equation}
The Vlasov equation is a sufficient condition for stress-energy
conservation. The particle number current given by
\begin{equation}
\label{Nmu}
N^{\mu} = \int p^\mu f \, dV_p 
\end{equation}
is also conserved.

Now we impose spherical symmetry. We write the generic spherically symmetric
metric in polar-radial coordinates, where it takes the form
\begin{equation}
\label{metric}
ds^2=-\alpha^2(t,r) dt^2+a^2(t,r)dr^2
     +r^2(d\theta^2+\sin^2\theta d\varphi^2).
\end{equation}
In this coordinate choice there is a residual gauge freedom $t\to
t'(t)$, which changes the lapse $\alpha$. We fix this freedom by
setting $\alpha(t,\infty)=1$. The Einstein equations give the
following equations for the first derivatives of the metric
coefficients:
\begin{eqnarray}
\frac{\alpha_{,r}}{\alpha}&=&
\frac{a^2-1}{2r}+4\pi ra^2{T_r}^r, \label{alphar} \\
\frac{a_{,r}}{a}&=&-\frac{a^2-1}{2r}-4\pi ra^2{T_t}^t, 
\label{ar} \\
\frac{a_{,t}}{a}&=&4\pi ra^2{T_t}^r.
\label{at}
\end{eqnarray}
The fourth Einstein equation, involving ${T_\theta}^\theta$, is a
combination of derivatives of these three, and so is redundant.

As geometric observables, we define
the Hawking mass
\begin{equation}
\label{Hawkingmass}
M_H(t,r):={r\over 2}\left[1-a^{-2}(t,r)\right]
\end{equation}
and, for static solutions $(a_0,\alpha_0)$, following the notation in
the literature, the maximum compactness
\begin{equation}
\label{Gammadef}
\Gamma:=\max_r{2M_H(r)\over r}=1-\left[\max_r a_0(r)\right]^{-2}
\end{equation}
and the central lapse and, equivalently, central redshift
\begin{equation}
\label{Zcdef}
\alpha_c:=\alpha_0(0), \qquad Z_c:={1\over \alpha_c}-1
\end{equation}
(not to be confused with the dynamical variable $Z$ introduced
below). 

In spherical symmetry, the distribution function has to be of the form
$f(t,r,p_r,|p|)$, where $|p|^2:= g_{ij}p^i p^j$. As is well-known, we
can simplify the Vlasov equation using the fact that angular momentum
is a constant of motion. We denote its square by
\begin{equation}
F := r^2 |p_\parallel|^2 = r^4\left[(p^\theta)^2+\sin^2\theta\,(p^\varphi)^2\right],
\end{equation}
where $p_\parallel$ is the 3-momentum vector tangential to the
symmetry orbits of constant $r$. The Vlasov equation and stress-energy
tensor in terms of $f(t,r,p_r,F)$ are given, for example, in
\cite{critvlasov}, and we do not repeat them here. Instead we adopt
a new set of variables designed to make the limit of vanishing
particle mass $m$ transparent and to simplify the equations in that
limit.

\subsection{New variables}

To motivate the new variables, we note that the massless
Einstein-Vlasov system has an additional symmetry. Two massless point
particles with the same starting point and with initial 4-momenta
$p_\mu$ and $\lambda p_\mu$, for any $\lambda>0$, follow the same
worldline. As a consequence, if two Vlasov distributions $f$ and $f'$
are related by $f'(x^\mu,p_\mu)=\lambda^4f(x^\mu,\lambda p_\mu)$ we
see from (\ref{Tmunuf}) that they give rise to the same stress-energy
tensor $T_{\mu\nu}$, and so are consistent with the same metric. (This
invariance does not hold for massive particles because the rest mass
$m$ does not scale. Put differently, two massive particles with initial 4-momenta
$p_\mu$ and $\lambda p_\mu$ follow different trajectories.)

To exploit this scale-invariance of the massless case in practice, we
need to be able to integrate out a particle momentum component that is
also a constant of the motion. In spherical symmetry, the absolute
value $\sqrt{F}$ of the particle angular momentum is conserved along particle
trajectories, and so can play this role, even in the time-dependent
case. Hence we replace the radial momentum $p_r$ by
\begin{equation}
z:={p_r\over a\sqrt{F}}.
\end{equation}
(The factor $a$ has been introduced for convenience.) It is easy to
see that for massless particles this is invariant under rescaling
$p_\mu$. The Vlasov equation for $f(t,r,z,F)$ is then
\begin{equation}
\frac{\partial f}{\partial t}
+{\alpha z\over aZ}\frac{\partial f}{\partial r}
+
\left({\alpha\over r^3 a Z}-{\alpha_{,r} Z\over a}-{z a_{,t}\over a}\right)
\frac{\partial f}{\partial z}
=0,
\label{VlasovtrzF}
\end{equation}
where we have defined the shorthand
\begin{equation}
\label{Zdefmassive}
Z(r,z,F):=\sqrt{{m^2\over F}+z^2+{1\over r^2}}.
\end{equation}
The partial derivative $\partial/\partial F$ does not appear in the
Vlasov equation because $F$ is a constant of motion. Furthermore, $m$
and $F$ appear in the Vlasov equation only in the combination
$m^2/F$. Hence the limit $m=0$ is regular.

Parameterising the tangential momentum in terms of $F$ and the angle
$0\le\chi<2\pi$ in the plane of tangential momenta,
\begin{equation}
p^\theta=:{\sqrt{F}\over r^2}\cos\chi, \quad 
\sin\theta\,p^\varphi=:{\sqrt{F}\over r^2}\sin\chi,
\end{equation}
we have 
\begin{equation}
dp^\theta\wedge d(\sin\theta\,dp^\varphi)={dF\wedge d\chi\over 2r^4}.
\end{equation}
With this intermediate step, it is then easy to see that the momentum
space volume element $dV_p$ given in (\ref{dVp}) in can be written in
the new variables as
\begin{equation}
dV_p={dF\wedge dz \wedge d\chi\over 2r^2 Z}.
\end{equation}
With 
\begin{eqnarray}
\int_{\chi=0}^{2\pi}dV_p&=&{\pi dF \wedge dz \over r^2 Z}, \\
\int_{\chi=0}^{2\pi}\sin^2\chi\,dV_p&=&{\pi dF \wedge dz \over 2r^2 Z},
\end{eqnarray}
the non-vanishing components of the stress-energy tensor
are given by the integrals
\begin{eqnarray}
p:=T_r{}^r &=& \frac{\pi}{r^2} 
\int_0^\infty F\,dF \int_{-\infty}^\infty f\, {z^2\over Z}\,dz \ge 0, 
\label{Trrz} \\
\rho:=-T_t{}^t &=& \frac{\pi}{r^2} 
\int_0^\infty F\,dF \int_{-\infty}^\infty f\, Z\,dz \ge 0, 
\label{Tttz} \\
j:=T_t{}^r &=& - \frac{\pi\alpha }{ar^2}
\int_0^\infty F\,dF \int_{-\infty}^\infty f\, z\,dz,
\label{Ttrz} \\
p_T:={T_\theta}^\theta = {T_\varphi}^\varphi  &=&  \frac{\pi}{2r^4}
\int_0^\infty F\,dF \int_{-\infty}^\infty f\, {1\over Z}\,dz\ge 0.
\label{Tphiphiz}
\end{eqnarray}
We note that in the Einstein equations $m$ and $F$ appear only in the
combinations $m^2/F$ and $F\,dF$. Again the limit $m=0$ is regular. 

$\rho$, $p$, $p_T$ and $j$ are the energy density, radial and
tangential pressure and energy current observed by observers at
constant $r$. We assume that $f\ge 0$ and that $f$ behaves in such a
way that these integrals exist and are finite at every point in
spacetime. ${T_\mu}^\nu$ is conserved and satisfies the dominant and
strong energy conditions. It is easy to see that ${T_\mu}^\mu=0$, or
$\rho+p+2p_T=0$, if $m=0$. The Einstein equations (\ref{ar},\ref{at})
can be rewritten in terms of $M_H$, $\rho$ and $j$ in the
pseudo-Newtonian form
\begin{eqnarray}
\label{mr}
M_{H,r}&=&4\pi r^2\rho, \\
\label{mt}
M_{H,t}&=&4\pi r^2j.
\end{eqnarray}

\subsection{Static solutions}
\label{section:staticsolutions}

A static spacetime has an additional Killing vector
\begin{equation}
\xi:={\partial\over\partial t},
\end{equation}
and hence in our coordinates the metric takes the form
\begin{equation}
\label{staticmetric}
ds^2=-\alpha_0^2(r) dt^2+a_0^2(r)dr^2
     +r^2(d\theta^2+\sin^2\theta d\varphi^2).
\end{equation}
The Einstein equations are then
\begin{eqnarray}
{\alpha_0'\over \alpha_0}&=&{a_0^2-1\over2r}+4\pi ra^2p_0,
\label{staticalpha} \\
{a_0'\over a_0}&=&-{a_0^2-1\over2r}+4\pi ra^2\rho_0.
\label{statica}
\end{eqnarray}

In a static spacetime, the particles have an additional constant of motion
\begin{equation}
E:=-\xi^\mu p_\mu=- p_t=\alpha_0 \sqrt{F}Z.
\end{equation}
A self-consistent static solution can be obtained from the ansatz
\begin{equation}
\label{Jeans}
f(r,p_r,F)=h(E,F).
\end{equation}
Static solutions with this ansatz for $m>0$, including solutions with matter at
the centre of spherical symmetry (``non-shells'') and with multiple
(thick) shells separated by vacuum regions have been constructed in 
\cite{AndreassonRein2007}.

In Newtonian physics, this ansatz includes all possible static
solutions, a result known as Jean's theorem, but in GR the most
general self-consistent static solution of Einstein-Vlasov in
spherical symmetry has the form \cite{critvlasov}
\begin{equation}
\label{GRJeans}
f(r,p_r,F)=
\sum_n \theta(r-r^{(n)}_-)\;\theta(r^{(n)}_+-r)\;h^{(n)}(E,F),
\end{equation}
where the limits $r_\pm^{(n)}$ of the $n$-th (thick) shell depend on
$h^{(n)}$ and the total mass further inside.  In other words, in GR
there can be more than one potential well, labelled by $(n)$, each
with a different Vlasov distribution function $h^{(n)}(E,F)$. For
simplicity of notation only, in the following we focus on the case of
a single potential well. Multi-well solutions can be
constructed by surrounding an existing solution with an additional
matter shell further out, taking into account that just inside the
additional shell the spacetime is not Minkowski but Schwarzschild. 

Instead of using $E$, we write the Vlasov function of a static
spherically symmetric solution as
\begin{equation}
f(r,z,F)={k}(Q,F), 
\end{equation}
where 
\begin{equation}
\label{Qdef}
Q(r,z,F):={E^2\over F}=\alpha_0^2 Z^2
\end{equation}
plays a similar role to $E$ but is invariant under rescaling $p_\mu$
for massless particles. We also change the integration variables in
the stress-energy from $z$ and $F$ to $Q$ and $F$. This gives
the stress-energy components
\begin{eqnarray}
p_0&=&{\pi\over \alpha_0^2r^2} 
\int_0^\infty F\,dF \int_{U}^\infty {k}\,{v}\,dQ, \label{p0} \\
\rho_0&=&-{\pi\over \alpha_0^2r^2} 
\int_0^\infty F\,dF \int_{U}^\infty {k}{1\over {v}}\,dQ, \label{rho0}
\end{eqnarray}
where we have defined the shorthands
\begin{eqnarray}
U(r,F)&:=&\alpha_0^2\left({m^2\over F}+{1\over r^2}\right), \\
\label{crdef}
{v}(r,Q,F)&:=&{z\over Z} 
=\sqrt{1-{{U}\over Q}}
={{|p_r|\over a_0}\over\sqrt{{p_r^2\over a_0^2}+{F\over r^2}}}
\end{eqnarray}
for the integration limit and integral kernel. Physically, $U$ is an effective
potential for the radial particle motion in the sense that
\begin{equation}
z^2=\alpha_0^{-2}(Q-U),
\end{equation}
and hence $Q=U$ determines the radial turning point for all particles
with a given $Q$ and $F$. From the last equality in (\ref{crdef}) we
see that physically $\pm{v}$ is the radial particle speed, expressed
in units of the speed of light and measured by observers at constant
$r$.

\section{The massless case} 
\label{section:massless}

\subsection{The reduced system} 
\label{section:reduced}

We now restrict to the case $m=0$. The key observation is that $Z$
then becomes independent of $F$.  Hence $F$ no longer appears in the
Vlasov equation (\ref{VlasovtrzF}) at all. This has two key
consequences: first, the integrated Vlasov density
\begin{equation}
\bar f(t,r,z):=\int_0^\infty f(t,r,z,F)F\,dF
\end{equation}
obeys the same PDE as $f$ itself, namely 
\begin{equation}
\frac{\partial \bar f}{\partial t}
+{\alpha z\over aZ}\frac{\partial \bar f}{\partial r}
+
\left({\alpha\over r^3 a Z}-{\alpha_{,r} Z\over a}-{z a_{,t}\over a}\right)
\frac{\partial \bar f}{\partial z}
=0,
\label{VlasovtrzFbar}
\end{equation}
where now
\begin{equation}
\label{Zdefmassless}
Z(r,z)= \sqrt{z^2+{1\over r^2}}.
\end{equation}

Secondly, the integration limit $U$ and kernel $v$ in the double
integrals for the stress-energy tensor become independent of $F$, and
we can therefore change the order of integration and write
\begin{eqnarray}
p&=& \frac{\pi}{r^2} 
\int_{-\infty}^\infty \bar f\, {z^2\over Z}\,dz \ge 0, 
\label{Trrzbis} \\
\rho &=& -\frac{\pi}{r^2} 
\int_{-\infty}^\infty \bar f\, Z\,dz \le 0, 
\label{Tttzbis} \\
j &=& - \frac{\pi\alpha}{ar^2}
\int_{-\infty}^\infty \bar f\, z\,dz,
\label{Ttrzbis} \\ 
p_T&=&  \frac{\pi}{2r^4}
\int_{-\infty}^\infty \bar f\, {1\over Z}\,dz\ge 0.
\label{Tphiphizbis}
\end{eqnarray}

We now have a closed system of field equations for the unknowns
$a(t,r)$, $\alpha(t,r)$ and $\bar f(t,r,z)$. We shall call this the
reduced system. For any solution $a(t,r)$, $\alpha(t,r)$ and
$f(t,r,z,F)$ of the full system, there are infinitely many other
solutions $a(t,r)$, $\alpha(t,r)$ and $f'(t,r,z,F)$, all
corresponding to the same solution of the reduced system,
$\bar f(t,r,z)=\bar f'(t,r,z)$.

\subsection{Static solutions} 
\label{section:reducedstatic}

In the static massless case, we can interchange the momentum
integrations in the same way. The static Einstein equations become
\begin{eqnarray}
{\alpha_0'\over \alpha_0}&=&{a_0^2-1\over2r}+{4\pi^2 a_0^2\over r\alpha_0^2}
\int_{U}^\infty \bar {k}\,{v}\,dQ, 
\label{alphamassless}
\\
{a_0'\over a_0}&=&-{a_0^2-1\over2r}+{4\pi^2 a_0^2\over r\alpha_0^2} 
\int_{U}^\infty \bar {k}{1\over {v}}\,dQ,
\label{amassless}
\end{eqnarray}
where  we have defined
\begin{equation}
\bar {k}(Q):=\int_0^\infty {k}(Q,F)F\,dF,
\end{equation}
and where now
\begin{equation}
U(r)={\alpha_0^2\over r^2}.
\end{equation}
$Q=U(r)$ now gives the turning points of all particles with a given
conserved quantity $Q$, independently of $F$.

\begin{figure}
\includegraphics[scale=0.6, angle=0]{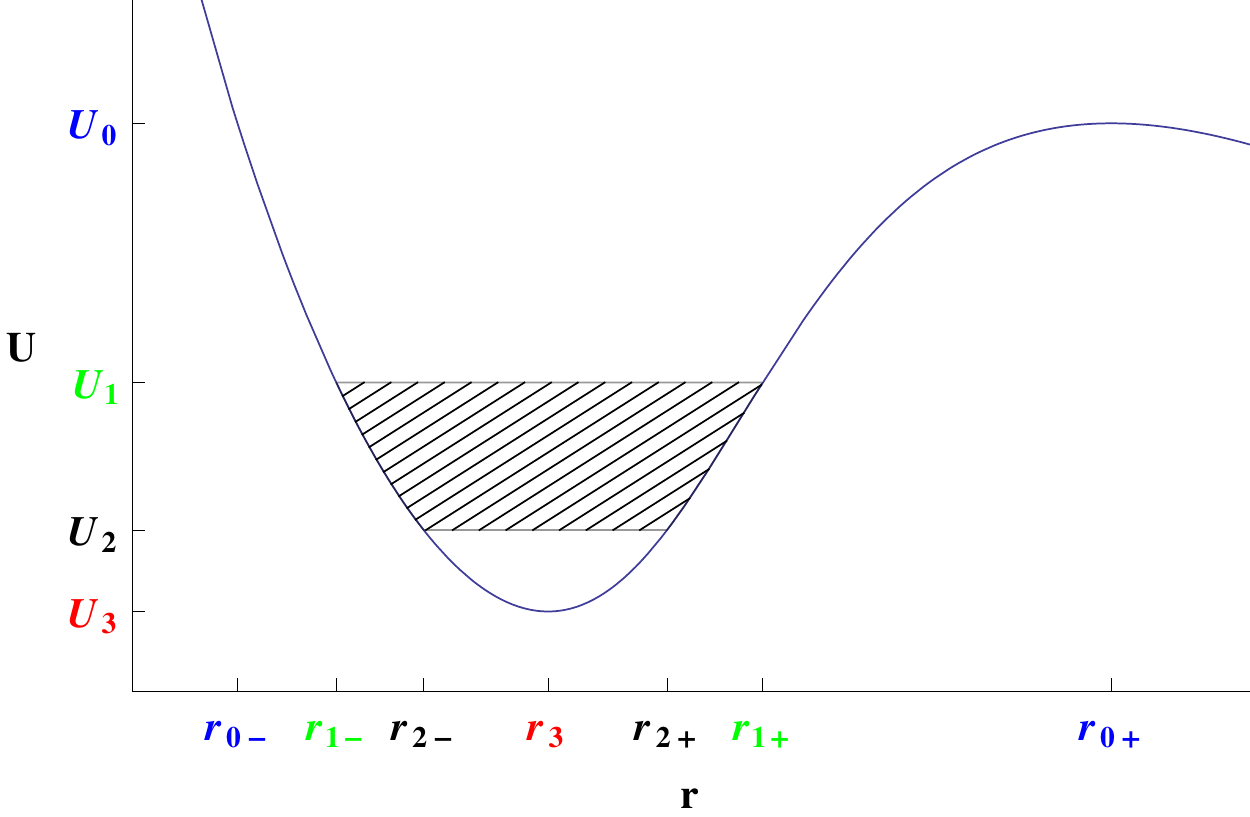} 
\caption{Sketch of the potential $U(r)$, showing the preferred values of
  $U$ and $r$ defined in (\ref{Uvals}-\ref{varphisupport}). The
  black hatching shows the region in $(U,r)$ space where particles are
  present.}
\label{figure:potentialsketch}
\end{figure}

For simplicity, we again assume that there is only a single potential
well in $U$. Throughout the remainder of the paper we will frequently
refer to the particular values of $r$ and $U$ illustrated in
Fig.~\ref{figure:potentialsketch}, and formally defined as follows:
\begin{eqnarray}
\label{Uvals}
U_3&\le& U_2\le U_1\le U_0, \\
r_{0-}\!&\le&\! r_{1-}\!\le r_{2-}\le r_3\le r_{2+}\le r_{1+}\le r_{0+}, \\
U(r_{i\pm})&=&U_i, \quad i=0,1,2, \qquad U(r_3)=U_3,\\
U'(r_3)&=&0, \qquad U''(r_3)>0, \\
U'(r_{0+})&=&0, \qquad U''(r_{0+})<0, \\
\bar k(Q)&\ne0 & \quad\Leftrightarrow\quad U_2<Q<U_1.
\label{varphisupport}
\end{eqnarray}
In words, $U_0$ is the top of the effective potential, $U_3$ its
bottom and $[U_2,U_1]$ is the support of $\bar k(Q)$, which for bound
particles must lie between the top and bottom of the potential. In the
limiting case $U_1=U_2$, $\bar k(Q)$ is a $\delta$-function, while
$U_1=U_0$ and $U_2=U_3$ correspond to the matter distribution filling
the potential well to the lip and to the bottom, respectively.

Fig.~\ref{figure:potentialsketch} also illustrates the following
observation. For $U_1<U_0$, the maximum $r_{0+}$ of the potential is
in vacuum, and even for $U_1=U_0$ it is still in vacuum from one
side. Hence it must be identical with the only maximum that the vacuum
potential has, also known as the unstable photon orbit of the
Schwarzschild solution. From this identification we read off
\begin{equation}
r_{0+}=3M, \qquad U_0={1\over 27 M^2}.
\end{equation}

For massless particles the Ricci scalar vanishes, and as a consequence
$a(t,r)$ and $\alpha(t,r)$ are related by an Einstein equation
in which $f$ does not appear. For static solutions,
this reduces to the ordinary differential equation (ODE)
\begin{equation}
\label{zeroRiccistatic}
{\alpha_0''\over\alpha_0}+{2\over r}\left({\alpha_0'\over\alpha_0}-{a_0'\over
  a_0}\right)
-{\alpha_0'\over\alpha_0}{a_0'\over a_0}+{1-a_0^2\over r^2}= 0.
\end{equation}
Replacing $a_0$ by the new dependent variable 
\begin{equation}
b(r):={1\over a_0^2},
\end{equation}
we see that (\ref{zeroRiccistatic}) is linear, inhomogeneous and first
order when considered as an ODE for $b$, namely
\begin{equation}
\left(r\alpha_0+{r^2\over 2}\alpha_0'\right)b'
+(r^2\alpha_0''+2r\alpha_0'+\alpha_0)b= \alpha_0.
\end{equation}
Given $\alpha_0$ and the regularity boundary condition $b(0)=1$, this
has a unique solution $b$, which moreover can be given in closed form
as an integral. At the same time, (\ref{zeroRiccistatic}) is linear,
homogeneous and second order when considered as an ODE for $\alpha_0$,
namely
\begin{equation}
\alpha_0''+\left({2\over r}+{b'\over 2b}\right)\alpha_0'
+\left({b'\over rb}+{b-1\over r^2}\right)\alpha_0= 0.
\end{equation}
Given $b$, the regularity boundary condition $\alpha_0'(0)=0$ and the
gauge boundary condition $\alpha_0(\infty)=1$, this has a unique
solution $\alpha_0$. For any trace-free matter, and in particular for
massless Einstein-Vlasov, $a_0(r)$ and $\alpha_0(r)$ are therefore
related one-to one.

As discussed before, we always normalise $t$ so that it is proper
time at infinity. Then in the interior vacuum region $[0,r_{1-})$,
\begin{equation}
\label{interiorvacuum}
\alpha_0(r)=\alpha_{\rm c}\le 1, \qquad a_0(r)=1, 
\end{equation}
while in the exterior vacuum region $(r_{1+},\infty)$,
\begin{equation}
\label{exteriorvacuum}
\alpha_0(r)=\left(1-{2M\over r}\right)^{1\over 2}, \qquad 
a_0(r)=\left(1-{2M\over r}\right)^{-{1\over2}},
\end{equation} 
where $M$ is the total (ADM) mass. 

\subsection{From metric to matter: Volterra equations} 
\label{section:Volterra}

We define the combination of metric derivatives
\begin{eqnarray}
\label{Ar}
A(r)&:=&\left({\alpha_0'\over\alpha_0}-{a_0^2-1\over 2r}\right){\alpha_0
  r^2\over 4\pi^2a_0^2}, \\
\label{Br}
B(r)&:=&\left({a_0'\over a_0}+{a_0^2-1\over 2r}\right){\alpha_0^3
  \over 4\pi^2a_0^2},
\end{eqnarray}
which are related to the stress-energy as
\begin{eqnarray}
\label{rhoB}
\rho_0&:=&{\pi\over r\alpha_0^3}B, \\
\label{pA}
p_0&:=&{\pi\over r^3\alpha_0}A.
\end{eqnarray}

In terms of $A$ and $B$, the Einstein equations
(\ref{alphamassless},\ref{amassless}) can be written as
\begin{eqnarray}
\label{Aru}
A(r)&=&\int_0^{u(r)}\sqrt{u(r)-q}\,\varphi(q)\,dq, \\
\label{Bru}
B(r)&=&\int_0^{u(r)}{\varphi(q)\over\sqrt{u(r)-q}} \,dq, 
\end{eqnarray}
where we have defined the new dependent and independent variables
\begin{equation}
\label{qvarphidef}
\varphi(q):=Q^2 \bar {k}(Q), \qquad q:={1\over Q}, 
\end{equation}
and the shorthand
\begin{equation}
\label{urdef}
u(r):={1\over U}={r^2\over\alpha_0^2}.
\end{equation}
It is easy to see from (\ref{Aru},\ref{Bru}) that 
\begin{equation}
B(r)=2{A'(r)\over u'(r)},
\end{equation}
and that via (\ref{Ar},\ref{Br}) this is equivalent to
(\ref{zeroRiccistatic}). 

The key observation is now that, from (\ref{Aru},\ref{Bru}), for any
two values of $r$ where $u(r)$ takes the same value, $A(r)$ must also
take the same value, and similarly for $B(r)$. Note this does not hold
in the massive case. On each of the two intervals $(r_{0-},r_3)$ and
$(r_3,r_{0+})$, $u(r)$ is monotonic. Using either one of these
intervals, we define the functions $\tilde A(u)$ and $\tilde B(u)$ on
the interval $(u_0,u_3)$ by
\begin{equation}
\label{ABABtilde}
\tilde A[u(r)]:=A(r), \qquad \tilde B[u(r)]:=B(r).  
\end{equation}
We have now turned
(\ref{alphamassless},\ref{amassless}) into linear Volterra equations
of the first kind, namely
\begin{eqnarray}
\label{Au}
\tilde A(u)&=&\int_0^u\sqrt{u-q}\,\varphi(q)\,dq, \\
\label{Bu}
\tilde B(u)&=&\int_0^u{\varphi(q)\over \sqrt{u-q}}\,dq,
\end{eqnarray}
From (\ref{Au},\ref{Bu}) they obey 
\begin{equation}
\label{A'B}
\tilde B(u)=2\tilde A'(u). 
\end{equation}

The integral equation (\ref{Bu}) is the Abel equation, which has the
unique solution \cite{encyclopedia,integralequations}
\begin{equation}
\label{Bphi}
\varphi(q)={1\over \pi}{d\over dq}\int_0^q
       {\tilde B(u)\over\sqrt{q-u}}\,du.
\end{equation}
The unique solution of the integral equation (\ref{Au}) is 
\cite{integralequations} 
\begin{equation}
\label{Aphi}
\varphi(q)={2\over \pi}{d^2\over dq^2}\int_0^q
       {\tilde A(u)\over\sqrt{q-u}}\,du.
\end{equation}
These two expressions for $\varphi(q)$ are equivalent modulo (\ref{A'B}).

In summary, we can uniquely obtain the matter from ``half the metric'' in
closed form as
follows.  With either $\alpha_0(r)$ or $a_0(r)$ given, we solve
(\ref{zeroRiccistatic}) for the other one. We obtain $A(r)$ and $B(r)$
by differentiation, and hence implicitly $\tilde A(u)$ and $\tilde
B(u)$. Finally, we obtain $\varphi(q)$ from either $\tilde A(u)$ or
$\tilde B(u)$ by integration and differentiation.

\subsection{From matter to metric: ODE boundary value problems} 
\label{section:shooting}

To obtain the metric from the matter, we first construct $\tilde A(u)$
and $\tilde B(u)$ by integration of $\varphi(q)$. We then consider
(\ref{Ar},\ref{Br}) with the identifications (\ref{ABABtilde}) as an
ODE system for $a_0(r)$ and $\alpha_0(r)$. For convenience, we change
the dependent variables from $\alpha_0$ to $u$ (because $u$ is the
argument of $\tilde A$ and $\tilde B$), and from $a_0$ to $b$ (because
the equations then become linear in $b$). With these changes,
(\ref{Ar},\ref{Br}) become
\begin{eqnarray}
\label{uprime}
rbu'+(1-3b)u+8\pi^2r^{-2}u^{3\over 2}\tilde A(u)&=&0, \\
\label{bprime}
rb'-1+b+8\pi^2r^{-2}u^{3\over 2}\tilde B(u)&=&0.
\end{eqnarray}

We now pose a boundary-value problem on the non-vacuum interval
$[r_{1-},r_{1+}]$, namely the ODE system (\ref{uprime},\ref{bprime})
with boundary conditions that arise from matching to Minkowski,
(\ref{interiorvacuum}), in the
interior and Schwarzschild, (\ref{exteriorvacuum}), in the exterior. These boundary conditions are
\begin{eqnarray}
\label{ur1m}
u(r_{1-})&=&u_1, \\
\label{br1m}
b(r_{1-})&=&1, \\
\label{ur1p}
u(r_{1+})&=&u_1, \\
\label{br1p}
b(r_{1+})&=&1-{2M\over r_{1+}}, 
\end{eqnarray}
where $r_{1-}$ is given by
\begin{equation}
\label{r1malphac}
r_{1-}=\alpha_c\sqrt{u_1},
\end{equation}
and $r_{1+}$ is the unique positive real root of the cubic
equation
\begin{equation}
\label{r1pcubic}
r_{1+}^3-u_1r_{1+}+2Mu_1=0.
\end{equation}
In the limit $u_1=u_0=27M^2$, we have $r_{1+}=r_{0+}=3M$ and
$b(r_{1+})=1/3$.

For given $\varphi(q)$, we now have a boundary value problem of two
first-order ODEs with two free parameters $M$ and $\alpha_c$. (Recall
that the value of $u_1$ is a property of the given function
$\varphi(q)$, namely, the upper limit of its support). Rather than
using $M$ as a parameter, however, we take advantage of the
scale-invariance of the massless Einstein-Vlasov equations (see
Appendix~\ref{appendix:scaleinvariance}) to fix $M=1$. We then use
$\alpha_c$ and an overall constant factor $C$ in $\varphi(q)$ as the
free parameters to be determined in the boundary value problem.

\section{Examples of static massless solutions} 
\label{section:examples}

\subsection{The thin-shell solution}
\label{appendix:thinshell}

We begin with the limiting case where all the matter forms a thin
shell at the bottom of the potential well. In this limit, $b$ becomes
discontinuous at the shell because the mass jumps. By contrast, $u$
must be continuous because the first Israel junction condition --- the
intrinsic metric is continuous across the shell --- is equivalent to
continuity of $r$ and $\alpha_0$. From the Einstein equations, we then
see that $u'$ is discontinuous.

In order to derive the thin-shell case as the formal limit of a family
of thick shells, we make the ansatz
\begin{eqnarray}
\label{bhatdef}
b(r)&=&\hat b(x), \\
\label{uhatdef}
u(r)&=&u_*+\epsilon \hat u(x), \qquad \hat u(0)=0
\end{eqnarray}
where 
\begin{equation}
\label{xdef}
x:={r-r_*\over\epsilon},
\end{equation}
which characterises a shell of width $\epsilon$ centered at
$r=r_*$. In the limit $\epsilon\to 0$, $b(r)$ jumps from $\hat
b(-\infty)$ to $\hat b(\infty)$, $u(r)$ is continuous with value
$u_*$, and $u'(r)$ jumps from $\hat u'(-\infty)$ to $\hat u'(\infty)$.

Taking the combination
$2\!\cdot\!(\ref{uprime})'-u'\!\cdot\!(\ref{bprime})$ in order to
cancel the diverging stress-energy term $\tilde B$ with $2\tilde A'$,
substituting the ansatz (\ref{bhatdef}-\ref{xdef}), and expanding to
leading order in $\epsilon$, we obtain that $\hat b\hat u'^2$ is
constant in $x$. This means that in the limit $\epsilon\to 0$ the
quantity $bu'^2$ is constant in $r$, and hence that it must take the
same value on either side of the thin matter shell. Matching $u$ and
$bu'^2$ from the interior vacuum (\ref{interiorvacuum}) to the
exterior vacuum (\ref{exteriorvacuum}) at some value $r_*$, we obtain
two algebraic equations that can be solved to find
\begin{equation}
r_*={9M\over 4}, \qquad \alpha_c={1\over 3}.
\end{equation}
and hence
\begin{equation}
u_*={729\over 16}M^2\simeq 45.6M^2, \qquad b_*=3.
\end{equation}
From $\alpha_c$ and $b_*$ (which is the global maximum of $b$) we have
the diagnostic quantities $Z_c=2$ and $\Gamma=8/9$. 

Note that we have not assumed anything about the function $\tilde
A(u)$, and so the thin shell limit is independent of the family of
thick shells from which it is obtained. It has already been shown
\cite{Andreasson2008} that the Buchdahl limit for (massless or
massive) static Einstein-Vlasov is $\Gamma\le 8/9$ and that the
thin-shell solution saturates it.

\subsection{Ansatz~1: $\varphi(q)$ a simple power}

A simple ansatz where we can transform between $\varphi$ and $\tilde
A$ or $\tilde B$ in closed form is
\begin{eqnarray}
\label{myvarphi}
\varphi(q)&=&{C}(k+1)u_1^{-1}\left[{q\over u_1}-1\right]_+^k, \\
\label{mytildeA}
\tilde A(u)&=&\tilde C u_1^{1/2}\left[{u\over u_1}-1\right]_+^{k+{3\over2}}, \\
\label{mytildeB}
\tilde B(u)&=&2\tilde A'(u)=\tilde C(2k+3)u_1^{-1/2}
\left[{u\over u_1}-1\right]_+^{k+{1\over2}},
\end{eqnarray}
with parameters $u_1\ge u_0$, $C$ and $k$. (Note $u_1$ has dimension
$L^2$, while $C$ is dimensionless, see
Appendix~\ref{appendix:scaleinvariance}.) We have introduced the
shorthand notation $[\dots]_+^k:=\theta(\dots)\,(\dots)^k$ and
\begin{equation}
\label{Ctildedef}
\tilde C:={\sqrt{\pi}\Gamma(k+2)\over 2\Gamma\left(k+{5\over
    2}\right)} C.
\end{equation}

In the limit $k\to-1$ we find
\begin{eqnarray}
\label{singleorbitvarphi}
\varphi(q)&=&C\delta\left({q\over u_1}-1\right), \\
\label{singleorbitAtilde}
\tilde A(u)&=&Cu_1^{1/2}\left[{u\over u_1}-1\right]_+^{1\over 2}, \\
\label{singleorbitBtilde}
\tilde B(u)&=&Cu_1^{-1/2}\left[{u\over u_1}-1\right]_+^{-{1\over2}},
\end{eqnarray}
and in this case $\tilde C=C$. [The factor $(k+1)$ has been included
  in the ansatz (\ref{myvarphi}) to make the limit $C\to 1$ regular.]
We shall call $k=-1$ the ``single-orbit'' case, as all particles now
follow the same trajectory (modulo orbital phase and orientation of
the orbital plane). As this single orbit is in general not circular,
the cloud of particles still forms a thick shell in physical space
$r$. (In the limit $u_1\to u_*$ this single orbit becomes
circular, and we again obtain the unique thin-shell solution.)

We now consider what range of $k$ is physical. From (\ref{rhoB}) and
(\ref{mytildeB}) we see that for $k>-1/2$, the energy density $\rho_0$
is continuous with the vacuum value of zero at the vacuum boundaries
$r=r_{1\pm}$, and for $k=-1/2$ it is discontinous but finite. Hence
for $k\ge -1/2$, the system (\ref{uprime},\ref{bprime}) is regular at
$u_1$. For $k<-1/2$, we must start off any numerical solution of
(\ref{uprime},\ref{bprime}) with an expansion about $u=u_1$. In
Appendix~\ref{appendix:u1expansions} we find such approximations for
$u_1>u_0$ and $-1\le k<-1/2$, but not for $u_1=u_0$ and $k\le
-1/2$. We conclude that the physical range of the parameters $k$ and
$u_1$ of our ansatz is given by
\begin{equation}
\label{physicalku1}
\begin{cases}
k\ge -1, & u_0< u_1<u_*, \\
k> -{1\over2}, &  u_1= u_0. 
\end{cases}
\end{equation}
(We have excluded $u_1=u_*$ here because we have already treated the
thin-shell case separately). From (\ref{pA}) and (\ref{mytildeA}) we
then see that, for any $k$, the radial pressure is continuous with the
vacuum value of zero and the tangential pressure $p_T$ approaches
$\rho$. Physically, this is so because particles move with the speed
of light, and at the vacuum boundary of a static solution by
definition that speed is tangential.
 
\subsection{Numerical examples of Ansatz 1}

We now describe numerical solutions of (\ref{uprime},\ref{bprime}) with
$\varphi(q)$ given by (\ref{myvarphi}). To have a good starting guess
for the shooting, we have explored the 2-dimensional parameter space
along 1-parameter families of solutions, keeping either $u_1$ or $k$
fixed.

A side benefit of using $C$ and $\alpha_c$ as free parameters is that
as long as the Einstein equations are regular at $r_{1\pm}$, we can
shoot from $r_{1+}$ given by (\ref{r1pcubic}) to decreasing $r$, and
adjust the single free parameter $C$ so that $u=u_1$ and $b=b_1$ occur
at the same $r$, which we identify as $r_{1-}$. $\alpha_c$ is obtained
from $r_{1-}$ by (\ref{r1malphac}). This means that we need to solve a
numerical root-finding problem only in the one parameter $C$.

For $k<-1/2$ we start off the integration into the matter region with
the approximate solutions for $u\gtrsim u_1$ given in
Appendix~\ref{appendix:u1expansions}, and shoot from both $r_{1-}$ and
$r_{1+}$ to a midpoint, which we choose as $r=r_*$. Our simple
implementation of the 2-parameter shooting, using the {\it
  Mathematica} functions {\tt NDSolve} and {\tt FindRoot} with their
default settings, is not very robust and fails near all three
boundaries of the half-strip (\ref{physicalku1}). Hence our coverage
of the physical parameter space (\ref{physicalku1}) is incomplete. We
improve on this by extrapolating our data points to the entire
$k\ge-1$, $u_0\le u_1\le u_*$ half-strip, which for plotting we
truncate at $k=4$ (there are no numerical problems at large $k$).

\begin{figure}
\includegraphics[scale=0.6, angle=0]{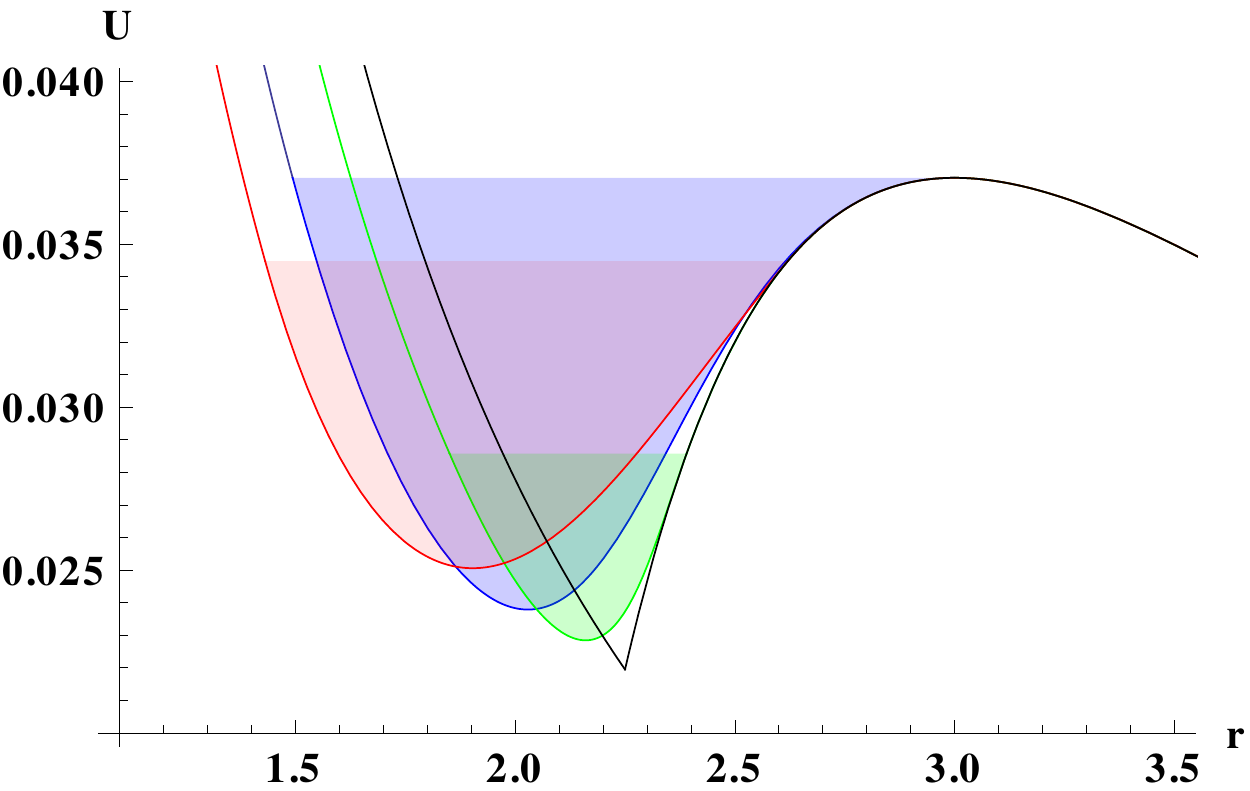} 
\includegraphics[scale=0.6, angle=0]{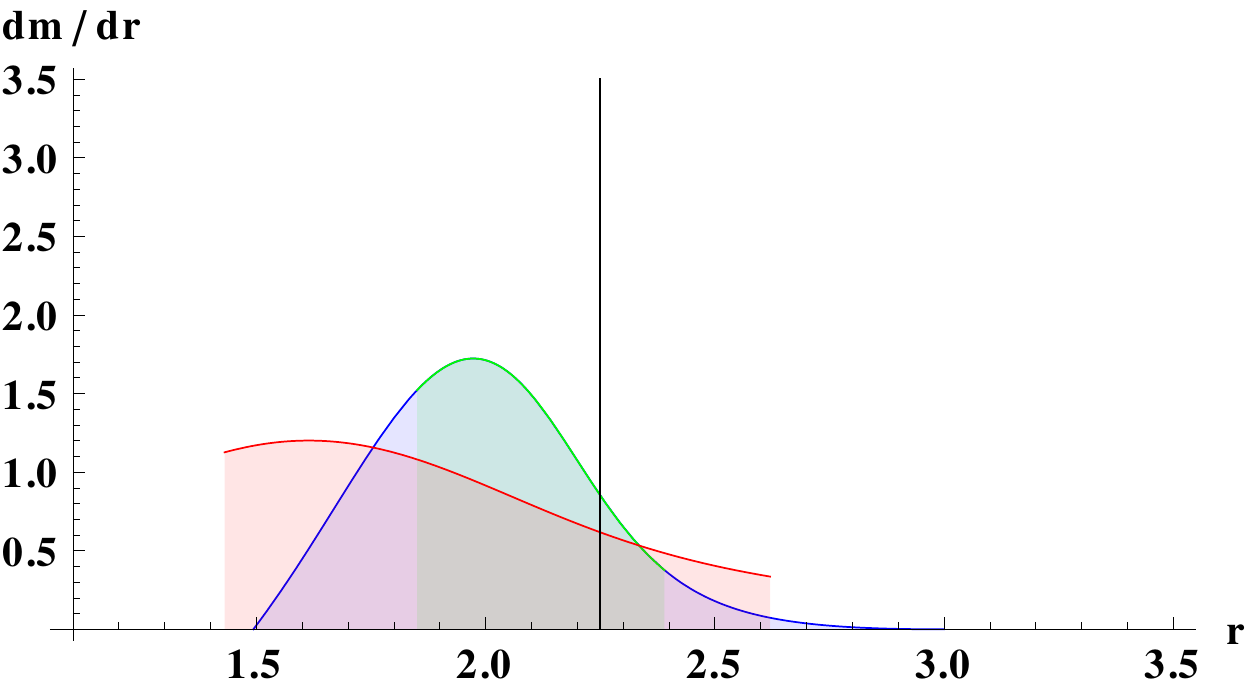} 
\caption{Effective potential $U(r)$ and mass aspect $dm/dr$ against
  radius $r$, for four representative solutions (from right to left in
  the upper plot, and from top to bottom in the lower plot): the
  thin-shell solution (black, $U'(r)$ discontinuous at $r=9/4$);
  $(k=1/2,u_1=35)$ (green); $(k=1/2,u_1=27)$ (blue, potential well
  completely filled); and $(k=-1/2,u_1=29)$ (red). Note that the area
  under $dm/dr$ is $M=1$ for all four solutions. $dm/dr=\delta(r-9/4)$
  has been symbolised by a vertical line. Note that the blue solution
  agrees with the critical solution of AC.}
\label{figure:representativepotentials}
\end{figure}

\begin{figure}
\includegraphics[scale=0.6,
  angle=0]{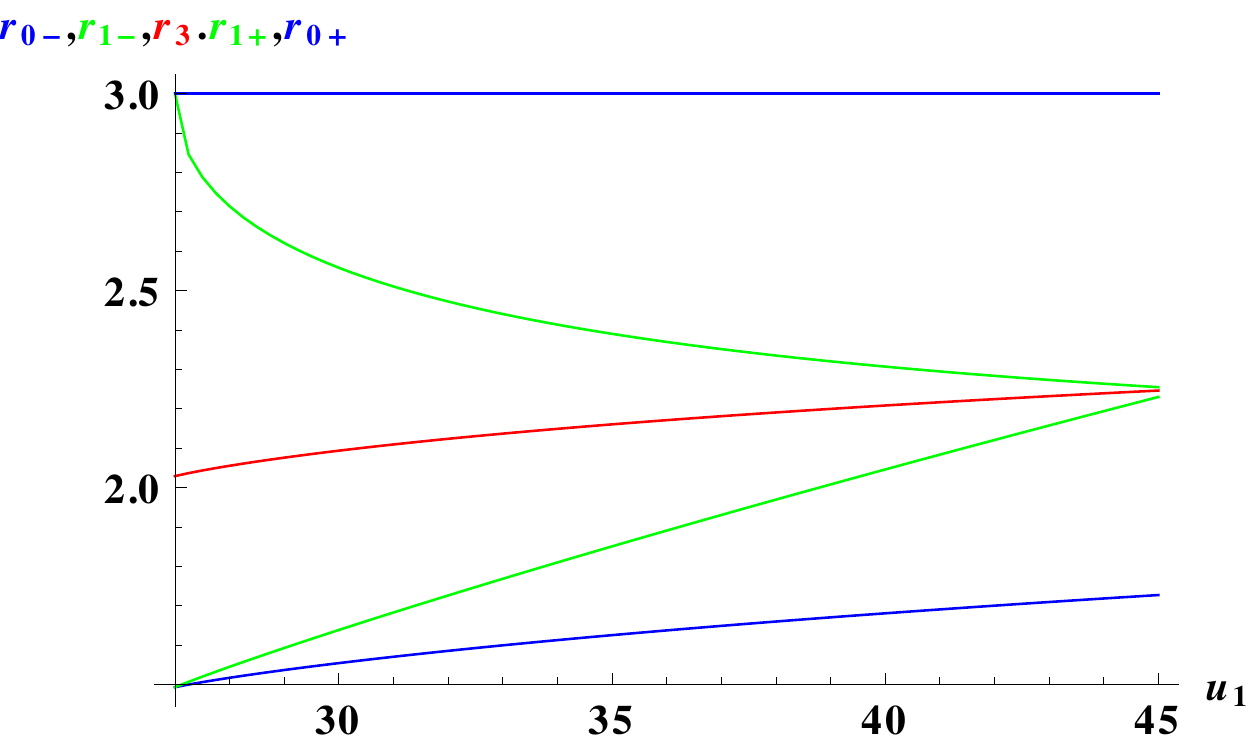}
\caption{From bottom to top, $r_{0-}$ (blue), $r_{1-}$ (green), $r_3$
  (red), $r_{1+}$ (green) and $r_{0+}=3$ (blue) against $u_1$ ranging from
  $u_0=27$ (the potential well is completely filled) to
  $u_*=729/16$ (the thin-shell solution), for Ansatz~1 with
  $k=1/2$. The colour-coding is the same as in
  Fig.~\ref{figure:potentialsketch}.}
\label{figure:ansatz1_keq1o2_locations}
\end{figure}

\begin{figure}
\includegraphics[scale=0.7, angle=0]{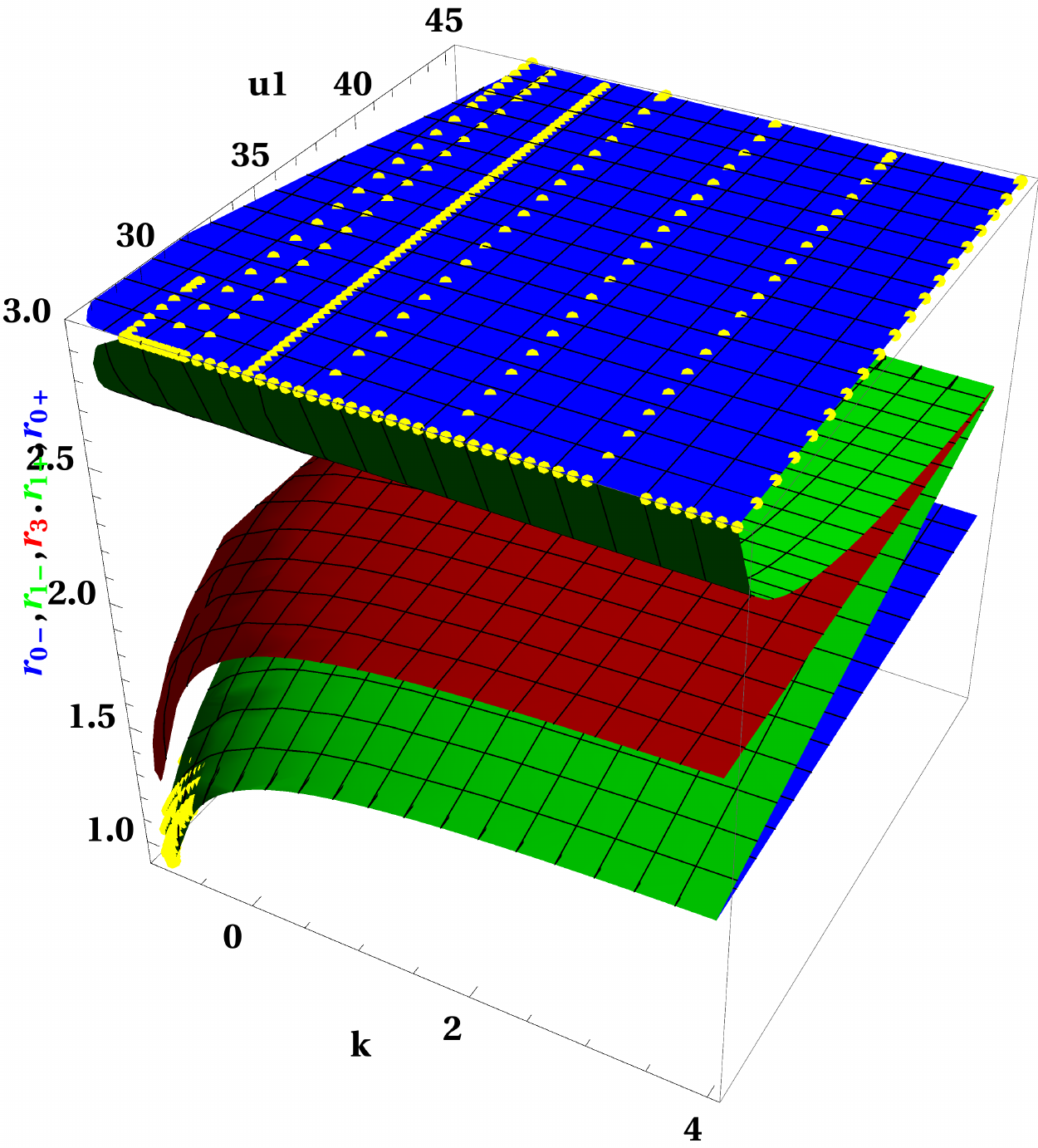}
\caption{From bottom to top, $r_{0-}$, $r_{1-}$, $r_3$, $r_{1+}$ and
  $r_{0+}=3$ against $k$ and $u_1$, for Ansatz~1. The colour coding is
  the same as in Figs.~\ref{figure:potentialsketch} and
  \ref{figure:ansatz1_keq1o2_locations}. The yellow beads
  indicate the discrete solutions we have computed, while surfaces
  represent interpolations. The plot ranges are $-1\le k\le 4$ and
  $u_0\le u_1\le u_*$. The top of the box is at $r_0=3$.}
\label{figure:ku1locations}
\end{figure}

\begin{figure}
\includegraphics[scale=0.7, angle=0]{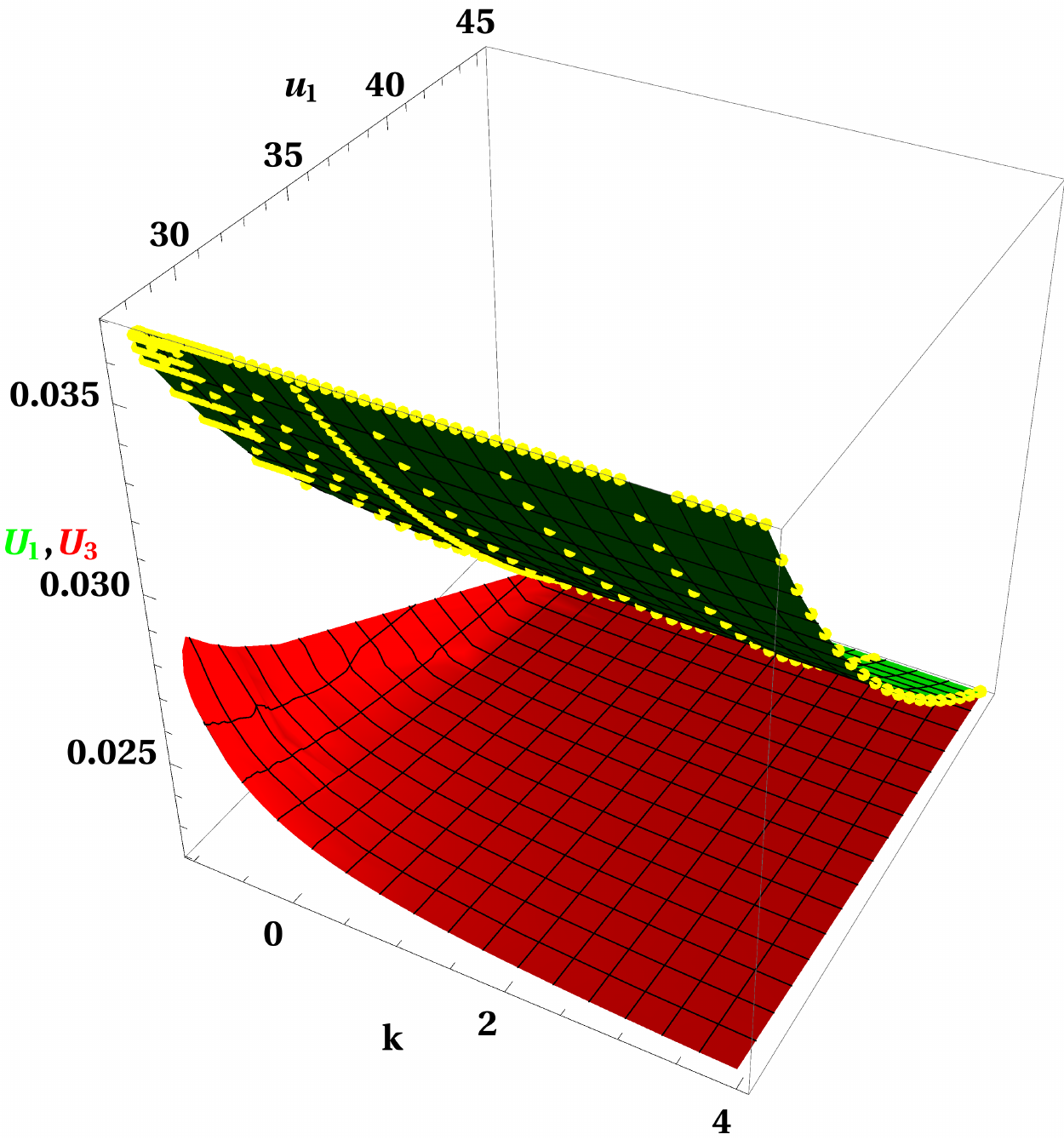}
\caption{$U_3$ (bottom) and $U_1$ (top) against $k$ and $u_1$, for
  Ansatz~1. The colour coding is the same as in
  Figs.~\ref{figure:potentialsketch},
  Fig.~\ref{figure:ansatz1_keq1o2_locations} and
  \ref{figure:ku1locations}. The top and bottom of the box are at
  $U_0=1/27$ and $U_*=729/16$, respectively. $U_3$ and $U_1$ come
  together at the value $U_*$ in the thin shell solution. In our family of
  solutions, this correspond to the limit $u_1\to u_*$, for any $k$, the back of
  the box.}
\label{figure:ku1U3U1}
\end{figure}

\begin{figure}
\includegraphics[scale=0.7, angle=0]{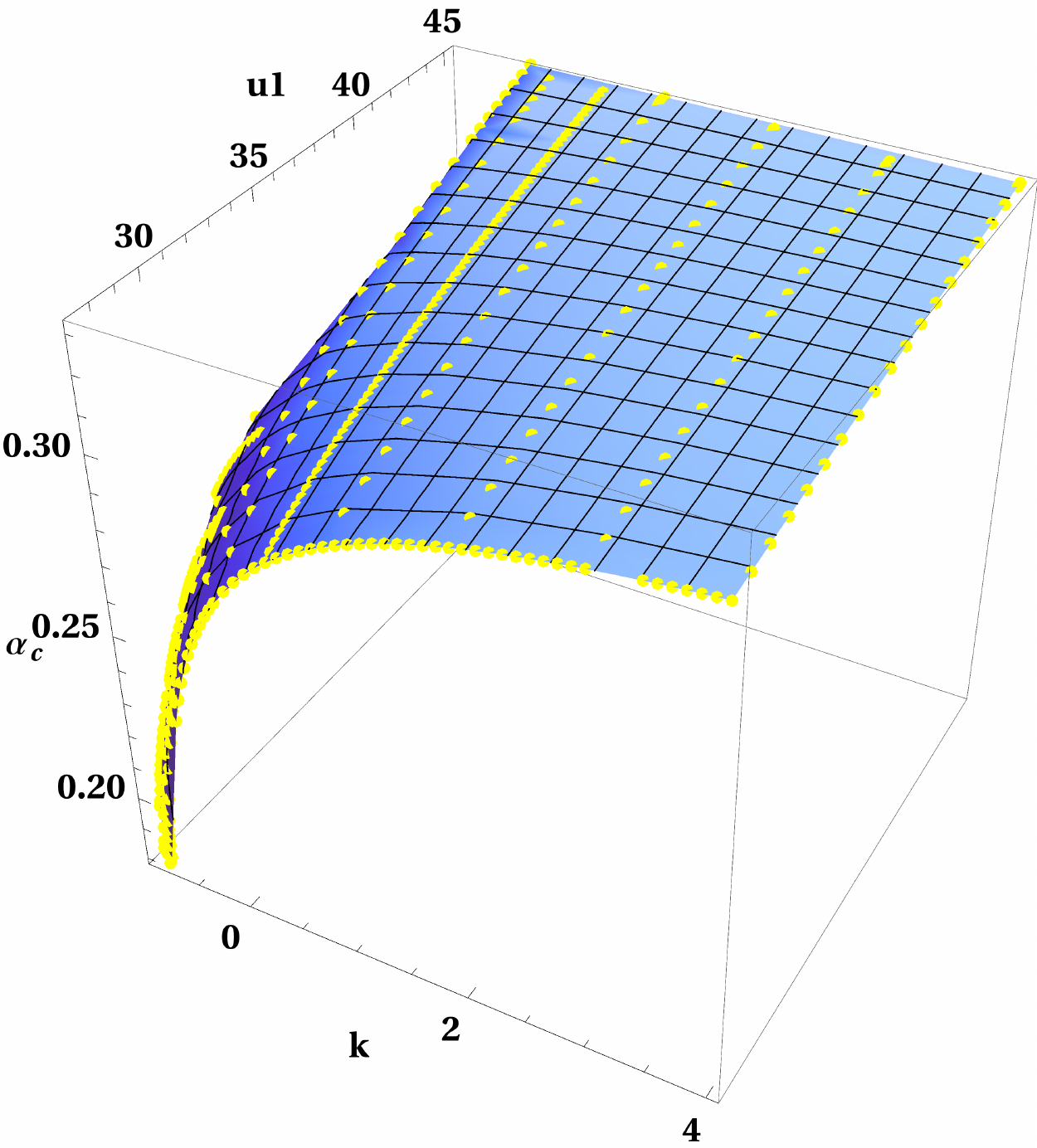}
\caption{$\alpha_c$ against $k$ and $u_1$, for Ansatz~1. The front of
  the box is at $u_0=27$, and the back at $u_*=729/16$. The top of
  the box is at $\alpha_{c*}=1/3$, the value for the thin shell
  solution. In our family of solutions, this corresponds to
  $u_1\to u_*$, for any $k$, the back of the box.}
\label{figure:ku1alphac}
\end{figure}

Fig.~\ref{figure:representativepotentials} shows the effective
potentials $U$ and mass aspect $dm/dr$ for four representative
solutions. Fig.~\ref{figure:ansatz1_keq1o2_locations} shows
$r_{1\pm}$, $r_{0\pm}$ and $r_3$ against $u_1$ for
$k=1/2$. Fig.~\ref{figure:ku1locations} shows the same locations, but
now against both $k$ and $u_1$, while Fig.~\ref{figure:ku1U3U1} shows
$U_3$ and $U_1$, and Fig.~\ref{figure:ku1alphac} the central lapse
$\alpha_c$, also against $k$ and $u_1$.

From these plots we see that the quantities $\Gamma$, $\alpha_c$,
$r_{0-}$, $r_{1\pm}$ and $r_3$ characterising the shape of the
potential well are all monotonic in both $k$ and $u_1$. In particular,
smaller $k$ or smaller $u_1$ give rise to less compact (smaller
$\Gamma$), more centrally redshifted (smaller $\alpha_c$) and more
spatially extended (smaller $r_{1-}$ and larger $r_{1+}$) solutions,
sitting in a wider potential well (smaller $r_{0-}$). (To be accurate,
$r_{1+}$ only depends on $u_1$.) By contrast, the most
compact, least spatially extended and least centrally redshifted
solution appears to be the unique thin-shell solution, which is
obtained in the limit $u_1\to u_*$, for any $k$.

\begin{figure}
\includegraphics[scale=0.6, angle=0]{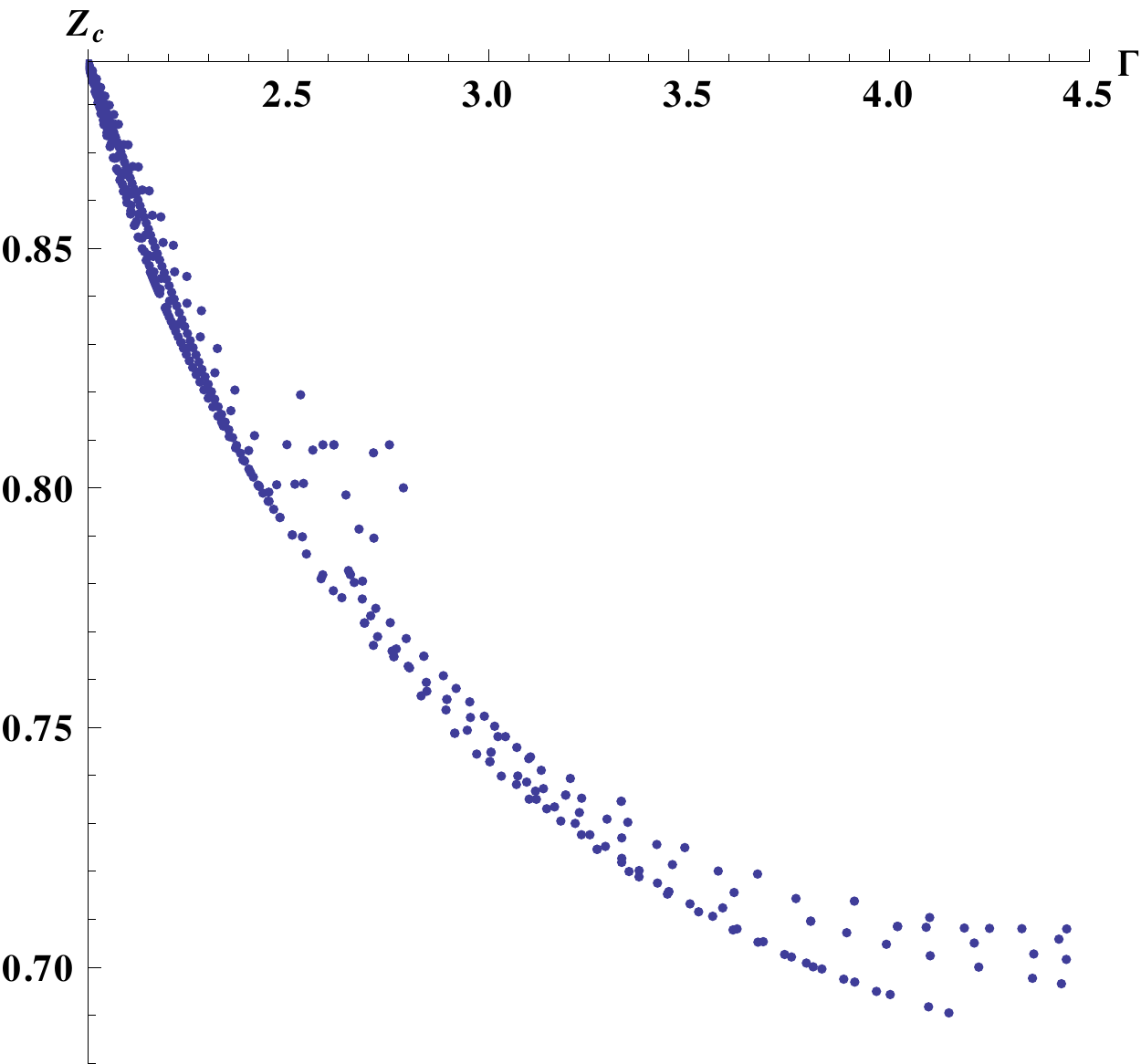} 
\caption{Projection into the $(Z_c,\Gamma)$-plane of the solutions from
  Ansatz~1 shown as yellow beads in Figs~\ref{figure:ku1locations} and
  \ref{figure:ku1U3U1}. As discussed in the text, these
  dots fill a narrow region. However, as that narrow region is
  ``almost'' the graph of a monotonic function, and as we have already
  plotted $\alpha_c$ against $(k,u_1)$, there is no need for us to
  plot $\Gamma$ against $(k,u_1)$. In this plot we show
  $Z_c:=\alpha_c^{-1}-1$, rather than $\alpha_c$, for direct
  comparison with Fig.~11 of AC.}
\label{figure:ansatz1_ZcGamma}
\end{figure}

Fig.~\ref{figure:ansatz1_ZcGamma} shows the projection into the
$(Z_c,\Gamma)$-plane of the data points shown as yellow dots in
Fig.~\ref{figure:ku1locations}. It is striking that these points lie
in a narrow strip, raising the question if there is in fact a
one-to-one relation between values of $Z_c$ and $\Gamma$, and the
deviations from it seen in this figure are due to numerical error.

Our own numerical results cannot settle this question, as we have not
attempted to quantify our numerical error. However, the equivalent
Fig.~11 of \cite{AkbarianChoptuik}, and in particular the inset of
that figure, shows that their 1-parameter families $k=1,2,3,4$ (in the
notation introduced in Appndix~\ref{appendix:ACansatz}) are distinct
curves in the $(Z_c,\Gamma)$ plane. They actually plot 2-parameter
families of ans\"atze $k(Q,F)$ that reduce to 1-parameter families
$\bar k(Q)$. The fact that these 2-parameter families clearly appear
as curves rather than areas in the $(Z_c,\Gamma)$ plane gives an
indication that their numerical error is much smaller than the
difference in the curves for different values of $k$. 

We can also make a, non-rigorous, theoretical argument that $Z_c$ is
not a function of $\Gamma$, using only
Eq.~(\ref{zeroRiccistatic}). For the sake of argument, consider
$\alpha_0$ as given, and $a_0$ obtained from it in closed
form. $\Gamma$ depending one-to-one on $Z_c$ would then be equivalent
to $\max_ra_0(r)$ depending only on $\alpha_0(0)$. Looking at the
explicit expression for $a_0$, this seems implausible.  Note that
because this argument only uses Eq.~(\ref{zeroRiccistatic}), it holds
for static spherically symmetric solutions with any tracefree
effective matter, often called ``geons''
\cite{AndreassonFajmanThaller}.

\section{The space of static solutions} 
\label{section:solutionspace}

Our mathematical results and numerical experiments suggest the
following three conjectures concerning the space of solutions. The
first two conjectures are an attempt at making more precise the notion
that ``locally'' the space of static spherically symmetric solutions
of the massless-Einstein Vlasov is a space of functions of one
variable, subject to certain integrability and positivity conditions.
The third conjecture is an attempt to characterise the subspace of
solutions that are analytic in the matter region. For definiteness, we
fix $M=1$. 

\begin{enumerate}

\item \label{conjectureI} For every function $\varphi(q)$ defined on the range
$[u_0,u_*]:=[27,729/16]$, such that 

\begin{enumerate}

\item \label{varphinontrivial} $\varphi$ is non-negative and not
  identically zero;

\item \label{varphintegrability} for all $u$ in this range the
  integrals $\tilde A(u)$ and $\tilde B(u)$ defined in
  (\ref{Au},\ref{Bu}) exist;

\item \label{Aufinite} $\tilde A(u)$ is finite;

\item \label{Bufinite} $\tilde B(u)$ is either finite or it is
  singular at isolated points not including $u_0$ and $u_*$, such that
  $\tilde B(u)$ is integrable;

\end{enumerate}

there exists a unique $C>0$ such that $C\varphi(q)$ creates a
self-consistent solution of (\ref{uprime},\ref{bprime}) with $M=1$ and
a single potential well.

\item \label{conjectureII} Conversely, for every static spherically
  symmetric metric of the form (\ref{staticmetric}) with $M=1$, such that

\begin{enumerate}

\item \label{metricnontrivial} $a_0$ and $\alpha_0$ are given by
  (\ref{interiorvacuum}) with $0<\alpha_c<1$ for $0\le r<r_{1-}$, and
  by (\ref{exteriorvacuum}) with $M=1$ for $r>r_{1+}$, where $0<r_{1-}< r_{1+}\le
  r_{0+}:=3$;

\item \label{metricdifferentiability} for $r_{1-}\le r\le r_{1+}$
  $A(r)$ and $B(r)$ defined by (\ref{Ar},\ref{Br}) exist and are
  non-negative;

\item \label{Arfinite} $A(r)$ is finite;

\item \label{Brfinite} $B(r)$ is either finite or it is singular at
  isolated points not including $r_{0+}$ such that $B(r)$ is
  integrable;

\item $a_0$ is $C^1$ and $\alpha_0$ is $C^2$ except at isolated
  points, and the two functions obey (\ref{zeroRiccistatic}) for
  $r_{1-}\le r\le r_{1}$ (in the sense of distributions at those
  isolated points);

\item \label{singlewell} $u(r)$ defined by (\ref{urdef}) is monotonically
  decreasing for $r_{1-}<r<r_3$ and monotonically increasing for
  $r_3<r<r_{1}$ for some $r_3$;

\item \label{a0range} $1\le a_0(r)<3$; 

\end{enumerate}

there exists a consistent solution with this metric and a matter distribution
$\varphi(q)$ given by (\ref{Bphi}) or (\ref{Aphi}). 

\item \label{conjectureIII} In particular, a class of solutions where
  the metric is analytic in $r$ in the non-vacuum region $r_{1-}\le
  r\le r_{1+}$ is obtained from the ansatz
\begin{equation}
\label{varphiseries}
\varphi(q)=\sum_{n=0}^\infty C_n \left[{q\over u_1}-1\right]_+^{n-{1\over2}},
\end{equation}
provided the series converges uniformly in $u_*\le q\le u_1$, and for
either

\begin{enumerate} 

\item $u_*<u_1<u_0$; or 

\item $u_1=u_0$ and $C_0=0$,

\end{enumerate}

If $C_{n_0}$ is the lowest non-vanishing coefficient in
(\ref{varphiseries}), the corresponding stress-energy tensor is given
to leading order near the inner vacuum boundary by
\begin{eqnarray}
\label{rholeading}
\rho(r)&\sim &[r-r_{1-}]_+^{n_0}, \\
p(r)&\sim&[r-r_{1-}]_+^{{n_0}+1}.
\end{eqnarray}
Similarly, in a one-sided neighbourhood of the outer vacuum boundary,
the leading behaviour is 
\begin{eqnarray}
\rho(r)&\sim &[r_{1+}-r]_+^{n_0}, \\
p(r)&\sim&[r_{1+}-r]_+^{{n_0}+1},
\end{eqnarray}
for $u_1>u_0$ (when $u-u_1 \sim r-r_{1+}$) and
\begin{eqnarray}
\rho(r)&\sim &[r_{1+}-r]_+^{2{n_0}}, \\
p(r)&\sim&[r_{1+}-r]_+^{2({n_0}+1)},
\label{pleading}
\end{eqnarray}
for $u_1=u_0$ [when $u-u_1 \sim (r-r_{1+})^2$].

\end{enumerate}

We now make a number of comments, in the order of the conjectures. 

\begin{enumerate}

\item For simplicity, we have excluded the thin-shell solution.

\item Conditions~\ref{varphinontrivial} and \ref{metricnontrivial} are
  related, each expressing that we have a shell solution with a vacuum
  centre. The existence of a vacuum centre, that is $r_{1-}>0$,
  follows from finite central redshift, that is (\ref{r1malphac}) with
  $\alpha_c>0$.

\item The integrability condition \ref{varphintegrability} and the
differentiability condition \ref{metricdifferentiability} are related,
expressing existence of the pressure and density.

\item Conditions~\ref{Aufinite} and \ref{Arfinite} are related, as are
  \ref{Bufinite} and \ref{Brfinite}, expressing positivity of the
  pressure and density, respectively.

\item Condition~\ref{Bufinite} is a straightforward generalisation of
  the case where $\rho$ diverges at $u_1$ for $u_1>u_0$ to a case
  where $u_1$ is replaced by $u_2$, or by an isolated singularity
  inside the nonvacuum region (using the linearity of the relation
  between $\rho$ and $\varphi$). A similar comment applies to
  condition~\ref{Brfinite}.

\item In Conjecture~\ref{conjectureII}, in a change of emphasis
  relative to Sec.~\ref{section:reducedstatic}, we have chosen to treat
  the metric as a single given entity obeying the trace-free
  constraint (\ref{zeroRiccistatic}), rather than splitting the
  conjecture into two, one taking $a_0$ as given and $\alpha_0$
  defined as the solution of (\ref{zeroRiccistatic}), and vice versa
  for the other. It is possible that Conditions~\ref{singlewell} and
  \ref{a0range} are redundant and can be derived from
  (\ref{zeroRiccistatic}).

\item Condition~\ref{singlewell} stipulates that there is a single
  potential well, and that it has a single minimum. As mentioned
  above, the extension to multishell solutions is straightforward because of
  Birkhoff's theorem. 

\item Condition \ref{a0range} is given by the positivity of the
  mass, and the fact that no other solution can be as compact as the
  thin-shell solution \cite{Andreasson2008}, where $a_0(r_*)=3$ at the
  shell. 

\item The ansatz (\ref{varphiseries}) is a linear superposition of
  examples of our Ansatz 1. The system (\ref{uprime},\ref{bprime}) of
  two first-order ODEs is then quasilinear with analytic coefficients,
  which suggests that solutions that remain regular are in fact
  analytic. (\ref{rholeading}-\ref{pleading}) follow from
  (\ref{mytildeA},\ref{mytildeB},\ref{rhoB}-\ref{pA}).

\item An obvious sufficient condition for the convergence of
  (\ref{varphiseries}) is that there exist an $N>0$, $K>0$ and
  $R>u_*-u_1$ such that $0\le C_n\le KR^{-n}$ for all $n>N$.

\end{enumerate}

\section{The critical solution}
\label{section:criticalsolution}

\begin{figure}
\includegraphics[scale=0.6, angle=0]{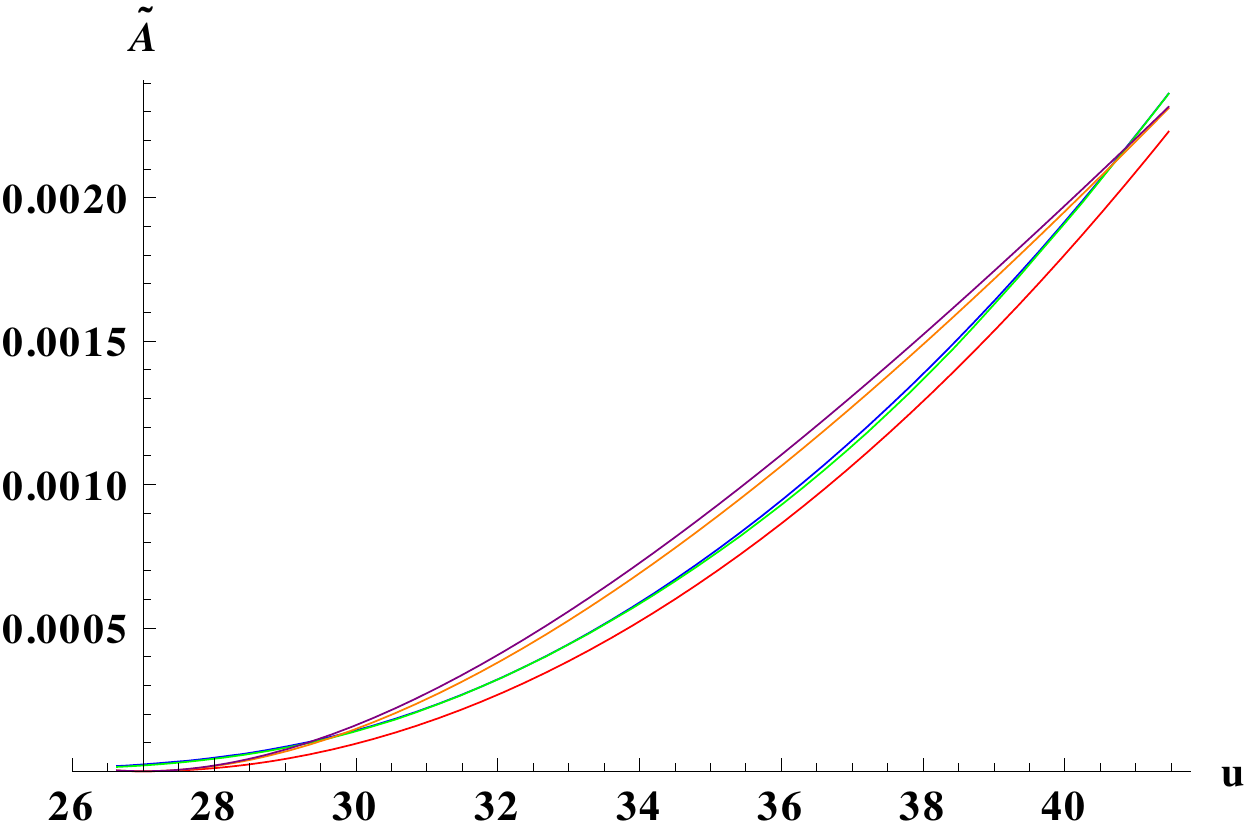} 
\includegraphics[scale=0.6, angle=0]{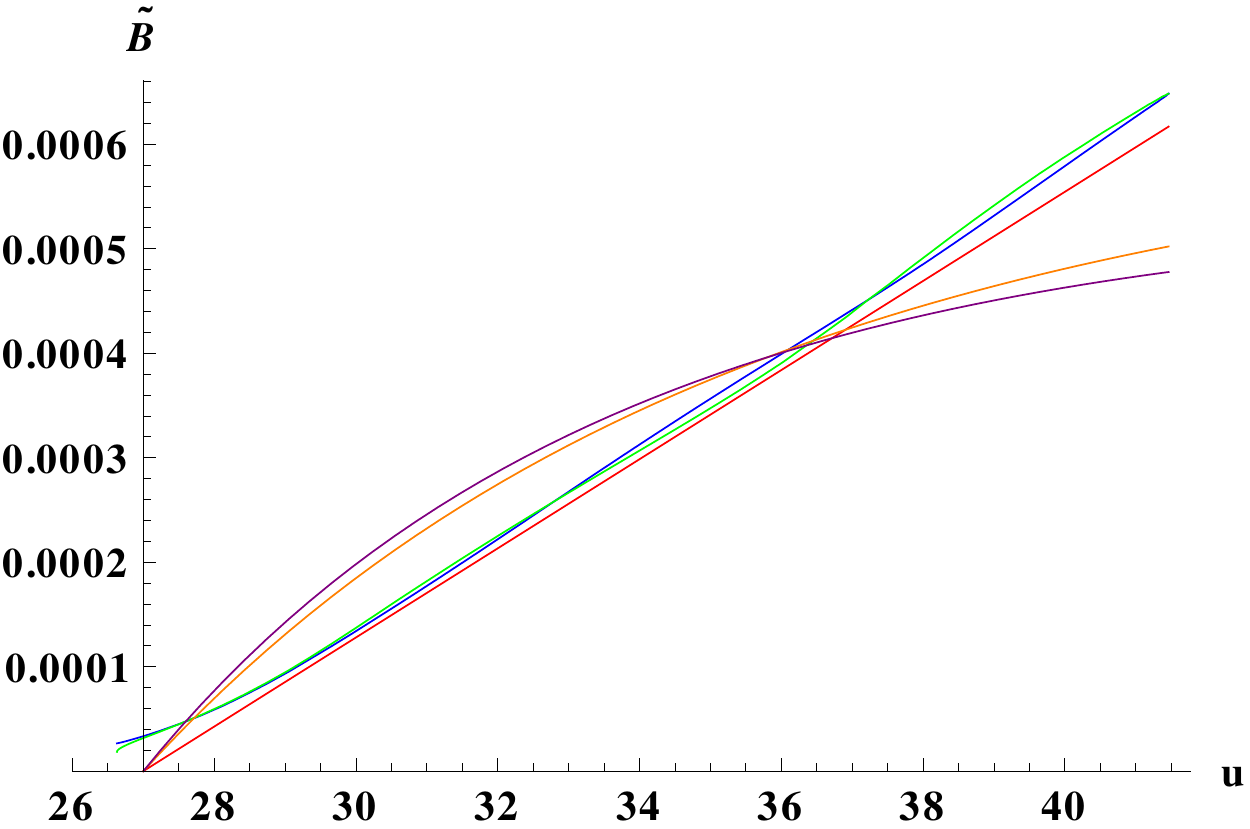} 
\caption{Plot of $\tilde A(u)$ (upper plot) and $\tilde B(u)$ (lower
  plot) inferred from the numerical values of $a_0(r)$ and
  $\alpha_0(r)$ corresponding to Figs.~7 and 8 of
  \cite{AkbarianChoptuik}, which represent the approximately universal
  critical solution. The three curves which lie almost on top of each
  other, correspond to the numerical data for $r_{0-}<r<r_3$ (blue)
  and $r_3<r<r_{0+}$ (green), and Ansatz~1 with $k=1/2$ and $u_1=u_0$
  (red). The other two curves correspond to Ansatz~2 with $k=1/2$ and
  $u_1=u_0$ (orange) and Ansatz~3 with $k=0$, $l=-1/2$ and $u_1=u_0$
  (purple). The axes origin is at $(27,0)$. Note that a
  self-consistent solution cannot extend below $u=27$ while the
  numerical curves do, indicating that the data of
  \cite{AkbarianChoptuik} are not exactly static, either because of
  limited fine-tuning or because of numerical error.}
\label{figure:AuBuplot}
\end{figure}

\begin{figure}
\includegraphics[scale=0.6, angle=0]{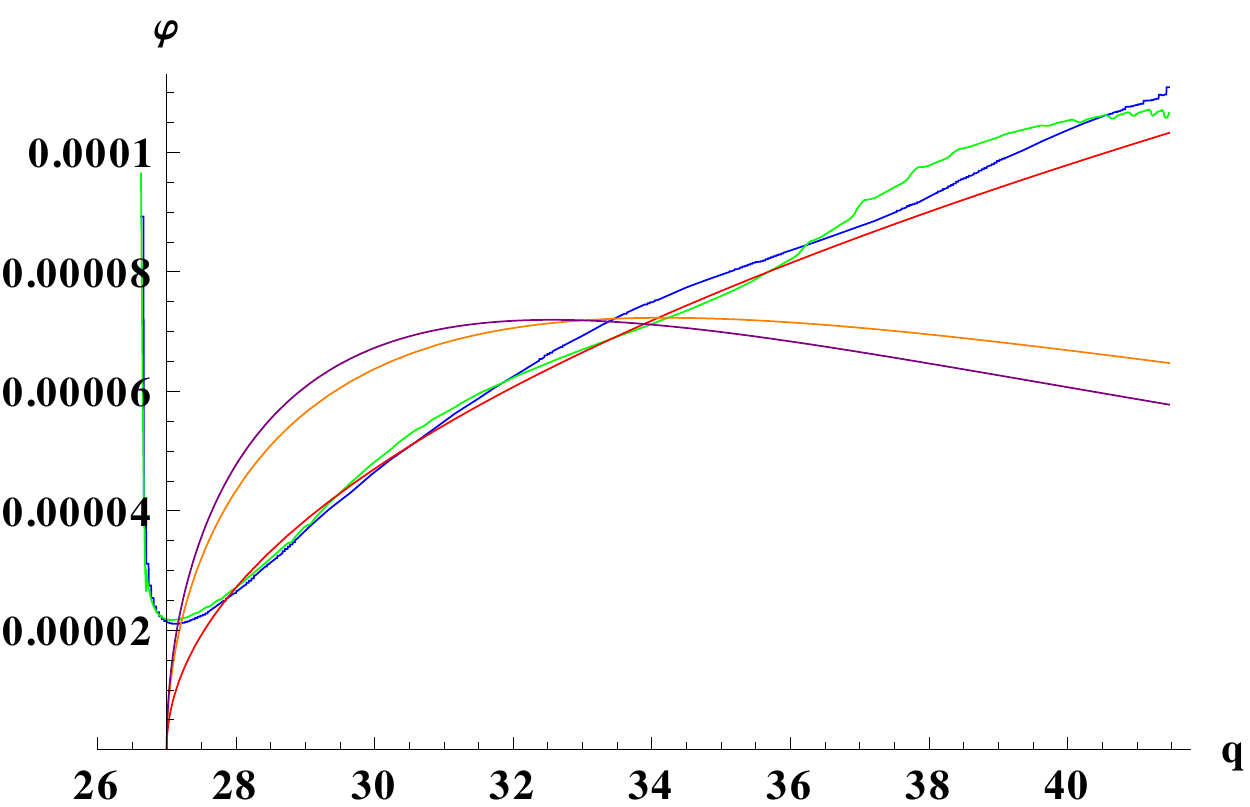} 
\caption{Value of $\varphi(q)$ obtained from the critical metric of
  \cite{AkbarianChoptuik} via $\tilde B(u)$, using the formula
  (\ref{Bphi}), and three exact solutions. The colour-coding of is the
  same as in Fig.~\ref{figure:AuBuplot}. From this and the previous
  plot, it is clear that Ansatz~1 is a good approximation to the
  critical solution (at least for the families of initial data
  considered by AC), while Ansatz~2 and 3 are not. Again, the critical
  solution found in time evolutions extends below $u=27$, meaning that
  it cannot be exactly static.}
\label{figure:Buplotbis}
\end{figure}

AC have kindly given us tabulated data for $a_0(r)$
and $\alpha_0(r)$ for a representative of their approximately
universal type-I critical solution obtained by fine-tuning generic initial
data, corresponding to their Figs.~7 and 8. The solution resulting
from the ansatz (\ref{myvarphi}) with $u_1=u_0$ and $k=1/2$,
corresponding to $\tilde B(u)\propto u-u_0$ and $\tilde A(u)\propto
(u-u_0)^2$, is a good fit, to within the error of their numerical data
(mostly a deviation from staticity). This is shown in
Figs.~\ref{figure:AuBuplot} and \ref{figure:Buplotbis}.

In \cite{critvlasov}, we noticed the degeneracy of the massless case
for both static and continuously self-similar solutions, and
conjectured it for their linear perturbations, but we overlooked that
it holds in general, as shown here in Sec.~\ref{section:massless}. We
went on to conclude that there could be no type-II critical phenomena
because any critical solution would have an infinite number of growing
modes $\delta f$ (but giving the same $\delta\bar f$) if it had
one. As AC correctly noted, this argument, if correct, would equally hold for
type-I critical phenomena. However, there is no such argument within
the reduced system. It remains entirely possible that the reduced
system admits solutions that are 1-mode unstable (with a unique
$\delta\bar f$).

In particular, if there exists a single 1-mode unstable static
solution, this would explain Result~(I) of AC, that at the black hole
threshold there is a type-I critical solution with universal metric
$(a_0,\alpha_0)$ up to an overall scale and universal critical
exponent, but family-dependent $f$. (However, $\bar f$ would then also
be universal up to scale.)

There is a tension between this and Result~(II) of AC, that every
static solution is at the threshold of collapse. Elsewhere in critical
collapse, the critical solution is uniquely identified by being 1-mode
unstable and at the threshold of collapse. If there is a continuum of
solutions obeying these two criteria, what singles out the universal,
or approximately universal, critical solution of AC's Result~(I) among
them?

We conjecture the following resolution of this apparent paradox.

\begin{enumerate}

\item Assume that in a near-critical time-evolution, a static critical
  solution is approached as the effective potential $U(t,r)$ becomes
  time-independent and any particles that are unbound in it evaporate
  to infinity. The potential well should then end up filled to the
  lip, or $U_1=U_0$ in our notation.

\item If the initial data are regular (for example smooth), we would
  expect the critical solution to be similarly regular, in particular
  at $U=U_1$. This selects a power-series solution of the form
  (\ref{varphiseries}) with $u_1=u_0$ and $C_0=0$. 

\item For initial data fine-tuned to the collapse threshold, but which
  are otherwise generic, all terms in (\ref{varphiseries}) should be
  present, starting in particular from $C_1\ne 0$. However, we have no
  argument for setting the higher $C_n$ in (\ref{varphiseries}). We
  conjecture that the critical solution is in fact only approximately
  universal.

\item There seems to be good agreement of $\varphi(q)\propto
  [q-27M^2]_+^{1/2}$ (Ansatz~1 with $k=1/2$ and $u_1=27M^2$) with the
  observed critical solution, as we demonstrated in
  Figs.~\ref{figure:AuBuplot} and \ref{figure:Buplotbis}. By
  comparison, Ansatz~2 with $k=0$, $l=-1/2$ and $u_1=u_0$, and Ansatz~3 with
  $k+l+1=1/2$ and $u_1=u_0$, both of which have the same leading power
  of $q-u_1$, but differ in higher powers, do not fit at all (see again
  Figs.~\ref{figure:AuBuplot} and \ref{figure:Buplotbis}). We can only
  guess that families of initial data sufficiently different from the
  ones evolved by AC would show a significantly different critical solution.

\end{enumerate}

\section{Conclusions} 
\label{section:conclusions}

We have reformulated the Einstein-Vlasov system in spherical symmetry
in a way in which the limit of vanishing particle mass $m$ is
transparent. We have used this to show that the space of massless
static solutions is essentially a space of functions of one
variable. This function can be taken to be any one of $a_0(r)$,
$\alpha_0(r)$, or $\varphi(q)$. Moreover, given either one of the
metric coefficients, we can give $\varphi(q)$ in closed
form. Conversely, $\varphi(q)$ determines the two metric coefficients
through an ODE boundary value problem. Modulo the conjecture that this
has a unique solution, there is then a one-to-one correspondence
between these different ways of characterising a static solutions. The
detailed conjecture is given Sec.~\ref{section:solutionspace}.

It is clear from our numerical examples that static spherically
symmetric solutions of Einstein-Vlasov with massive particles are
highly general-relativistic. An interesting open question is therefore what
solutions achieve the smallest inner radius of the matter shell, the
lowest maximal compactness $\Gamma$ and largest central redshift
$Z_c$. A priori this could be three different solutions, or families
of solutions, but our numerical plots suggest that these are achieved
by a single unique solution, given by the limit $k\to-1$ and $u_1\to
u_0$. (However, the limiting point $k=-1$, $u_1=u_0$ is not itself
part of the solution space).

By contrast, we already know that the most compact, least spatially
extended and least centrally redshifted solution is the unique
thin-shell solution, where the shell is at $r=9M/4$ and hence
$\Gamma=8/9$ and $Z_c=2$. In any case, by considering $k$ down to
$-1$, we have already explicitly constructed solutions with
compactness as low as $\Gamma\simeq 0.7$, whereas $AC$ only considered
$k\ge 1$ and \cite{AndreassonFajmanThaller} seem to have explored only
values of the equivalent parameter $k+l+1\ge 0.7$.

We have also gone part of the way towards understanding the recent
numerical results of Akbarian and Choptuik on the stability of static
solutions and type-I critical collapse in the massless spherically
symmetric Einstein-Vlasov system. We have shown that there is,
contrary to an incorrect argument implicit in \cite{critvlasov}, no
reason why static solutions cannot be 1-mode unstable, and so no
contradiction with their Result (II). In
Sec.~\ref{section:criticalsolution} we have attempted to also
reconcile their Results (I) and (II) with each other: we conjecture
that there is no unique critical solution, but that the
family-dependent critical solutions all look similar if the initial
data are sufficiently generic, in the sense that all allowed powers of
$q-u_*$ are present in $\varphi(q)$.

We hope that this non-universality can be confirmed by more accurate
numerical simulations of critical collapse, using a wider variety of
1-parameter families of initial data. Our reformulation of the
massless system in terms of the dependent variable $\bar f$ and
independent variables $(t,r,z)$ should help with this in two
ways: Reducing the number of independent variables of the problem (not
counting time) from 3 to 2 allows a significant increase in resolution
without loss of generality, while separating trivial from nontrivial
parameters in the initial data allows fuller exploration of the space of
generic initial data.

Clearly, in order to fully understand type-I critical phenomena, we need to
obtain the perturbation spectrum of static solutions. In particular,
we would like to identify the apparently unique unstable mode of all static
solutions and understand its apparent universality. 

We note for possible future use that in axisymmetry the angular
momentum component $L_z$ is still conserved, and so we can adapt the
formalism introduced here to reduce the axisymmetric massless
Einstein-Vlasov system by one independent variable, from
$f(t,\rho,z,p_\rho,p_z,L_z)$ to $\bar f(t,\rho,z,p_\rho/L_z,p_z/L_z)$,
where, in this expression only, $\rho,z,\varphi$ are the standard
cylindrical coordinates. This means that the number of independent
variables (not counting time) reduces from 5 to 4. 

\acknowledgements

I am grateful to H\aa kan Andr\'easson and Simone Calogero for
inviting me to G\"oteborg, to H\aa kan, Maximilian Thaller and Ellery
Ames for helpful and inspiring discussions that gave rise to this
paper, and to Chalmers University of Technology for financial
support. I am also grateful to Arman Akbarian for a discussion of his
work and for making available numerical data relating to the critical
solution, and to Matt Choptuik for comments on a draft.

\appendix

\section{Dimensional analysis}
\label{appendix:scaleinvariance}

In gravitational units where $c=G=1$, all geometric quantities can be
assumed to have dimensions that are powers of length $L$. In particular,
we can assume that
\begin{equation}
[{T_\mu}^\nu]=L^{-2}, \quad [a]=[\alpha]=1, \quad [t]=[r]=[M]=L,
\end{equation}
where the metric is given by (\ref{metric}).
In Einstein-Vlasov, it is both consistent and convenient for
dimensional analysis to formally retain a separate dimension $P$ of
particle momentum. For massive particles, momentum is particle mass
$m$ times velocity (which is dimensionless for $c=1$), so $m$ also has
dimension $P$. We then have
\begin{eqnarray}
&&[p_r]=[m]=[E]=P, \quad [F]=P^2L^2, \\
&&[f]=[k]=P^{-4}L^{-2}, \quad [dV_p]=P^2.
\end{eqnarray}
We have defined our new variables so that powers of $P$ cancel, namely
\begin{eqnarray}
&&[z]=[Z]=L^{-1}, \\
&&[U]=[Q]=L^{-2}, \quad [\bar f]=[\bar k]=L^2, \\
\label{uvarphiscaling}
&&[u]=[q]=L^2, \quad [\varphi]=L^{-2}, \\
&&[A]=L, \quad [B]=L^{-1}.
\end{eqnarray}

How particle momentum $P$ scales with $L$ is a matter of convention,
and has no physical significance in the context of the massless
Einstein-Vlasov system. However, if one insists on giving this
physical significance, one natural choice is that particle momentum
scales as spacetime momentum, and so $P=L$.  This corresponds to
increasing the mass and size of a static solution with massless
particles by using the same number of particles but scaling up their
momenta. Another natural choice is that linear particle momentum is
invariant under rescaling (because in the massive it would be natural
for $m$ to be invariant), and so $P=1$. This corresponds to scaling up
the mass and size of a static solution with massless particles by
using more particles of the same momentum.

\section{Expansion of the shooting equations about $u_1$}
\label{appendix:u1expansions}

With (\ref{mytildeA},\ref{mytildeB}), for $-1\le k<-1/2$, $u$
and $b$ are continous but $b'$ diverges at $r=r_{1\pm}$. Hence to
start up a numerical solution of (\ref{uprime},\ref{bprime}) we need
to expand the solution about these singular points. We make the ansatz
\begin{eqnarray}
\label{uapprox}
u(r)&\simeq&u_1+\alpha(r-r_1)^p, \\
\label{bapprox}
b(r)&\simeq&b_1+\beta(r-r_1)^q,
\end{eqnarray}
and retain only the leading power of $r-r_1$ in each of the three
terms in (\ref{uprime},\ref{bprime}). We can then attempt to match the
power and coefficient in two of these terms in each equation, thus
giving us four algebraic equations for $(\alpha,\beta,p,q)$, and then
verify that neglecting the remaining term in each equation is
self-consistent for the resulting values of $(p,q)$.
Neglecting the third term in both (\ref{uprime}) and (\ref{bprime})
gives the linearised vacuum Einstein equations, and so is not a
relevant approximation, leaving us with $3\cdot 3-1=8$ other possible
combinations of neglected terms. We need to distinguish three values
of $b_1$, as follows.

\paragraph*{Matching to the interior vacuum region} 

In this special case $b(r_{1-})=1$, and so we can approximate $1-3b\simeq -2$
in (\ref{uprime}). Neglecting the third term of (\ref{uprime}) and the
second term of (\ref{bprime}), we find 
\begin{eqnarray}
\label{ucase1}
u-u_1&\simeq& 2u_1{r-r_{1-}\over r_{1-}}, \\
\label{bcase1}
b-1&\simeq& -\hat C_-2^{k+3/2}\left({r-r_{1-}\over r_{1-}}\right)^{k+3/2},
\end{eqnarray}
where we have defined the shorthands
\begin{equation}
\hat C_\pm:={8\pi^2 u_1\over r_{1\pm}^2}C.
\end{equation}
The conditions for the two neglected terms to really be subdominant
reduce to $k>-3/2$. Any other choice of neglected terms is
inconsistent.

\paragraph*{Matching to the exterior vacuum region when the
potential is full} 

In this special case we have $r_{1+}=3$ and $u_1=27$ (as always
setting $M=1$) and $b(3)=1/3$. It is clear that for $k<-1/2$ the
second term of (\ref{bprime}), which is finite, can be neglected with
respect to the third term, which diverges as $u\to u_0$. If we neglect
the third term of (\ref{uprime}), we find consistent powers, but the
coefficients are complex. In the other two cases the powers are not
consistent. This suggests that there is no continuous solution with
$k<-1/2$ and $u_1=u_0$.

\paragraph*{Matching to the exterior vacuum region when the
potential is not completely filled}

In the generic case we have $b_1\ne 1,1/3$, and $r_{1+}$ and $b_1$ are
given in terms of $u_1$ by (\ref{br1p}) and
(\ref{r1pcubic}). Neglecting the third term of (\ref{uprime}) and the
second term of (\ref{bprime}), we find
\begin{eqnarray}
\label{ucase3}
u-u_1&\simeq&u_1{(1-3b_1)\over b_1}{r_{1+}-r\over r_{1+}}, \\
\label{bcase3}
b-b_1&\simeq&-2\hat C_+
\left({1-3b_1\over b_1}\right)^{k+1/2}
\left({r_{1+}-r\over r_{1+}}\right)^{k+3/2}.
\end{eqnarray}
This is consistent for $-3/2<k<-1/2$ and the other two possibilities are not.

\section{Ansatz~2: Akbarian and Choptuik}
\label{appendix:ACansatz}

To obtain solutions that are static by construction, AC make the ansatz 
\begin{equation}
\label{ACstatic}
h(E,F)=C_{\rm AC}\left[1-{E\over E_0}\right]^k_+\delta(F-F_0).
\end{equation}
In our formalism this corresponds to
\begin{eqnarray}
\label{ACvarphi}
\varphi(q)&=&C\,2^k u_1^{-1}  \left({u_1\over
  q}\right)^2\left[1-\sqrt{u_1\over q}\right]_+^k \\
\label{ACvarphiapprox}
&\simeq&C(k+1)u_1^{-1}\left[{q\over u_1}-1\right]_+^k,
\end{eqnarray}
with the parameters in (\ref{ACvarphi}) related to those in
(\ref{ACstatic}) by
\begin{eqnarray}
\label{u1E0F0}
u_1&=&{F_0\over E_0^2}, \\
\label{CAC}
C&=&2^{-k}(k+1)^{-1}E_0^2C_{\rm AC}.
\end{eqnarray}
Note that $C_{\rm AC}$ has dimension. We have normalised $C$ so that
(\ref{ACvarphi}) agrees to leading order with (\ref{myvarphi}) for the
same $C$.  The ansatz (\ref{ACvarphi}) can be integrated in closed
form for $k>-1$ to give
\begin{eqnarray}
\tilde A(u)&=&\tilde C u_1^{1/2}
\left[{u\over u_1}-1\right]_+^{k+{3\over 2}} \left({u_1\over u}\right)^{k+1}\nonumber\\ &&
{}_2F_1\left({k+1\over 2},{k+2\over2},{5\over k+1},1-{u_1\over
  u}\right)
\nonumber \\
&\simeq&\tilde Cu_1^{1/2}\left[{u\over u_1}-1\right]_+^{k+{3\over 2}}, \\
\tilde B(u)&=&2\tilde A'(u)\simeq \tilde C(2k+3)u_1^{-1/2}\left[{u\over u_1}-1\right]_+^{k+{1\over 2}},
\end{eqnarray}
with $\tilde C$ again defined by (\ref{Ctildedef}).
In the limit $k\to -1$ we again obtain the
single-orbit case (\ref{singleorbitvarphi}-\ref{singleorbitBtilde}). 

Akbarian and Choptuik noted that all their explicitly constructed
static solutions with the two parameters $F_0$ and $E_0$, for fixed
$k$, lie on a single curve in $(Z_c,\Gamma)$. This is expected because this
2-parameter family in terms of $k(Q,F)$ corresponds to a 1-parameter
family in terms of $\bar k(Q)$. 
 
\section{Ansatz~3: Andr\'easson, Fajman and Thaller}
\label{appendix:AFTansatz}

Andr\'easson, Fajman and Thaller \cite{AndreassonFajmanThaller} have
proved the existence of solutions for the ansatz 
\begin{equation}
\label{PhiAFT}
h(E,F)=[E_0-E]^k_+[F-F_0]_+^l
\end{equation}
with $l\ge -1/2$, $k\ge 0$, $F_0\ge 0$ and $E_0>0$. In our formalism
this corresponds to
\begin{eqnarray}
\label{varphiAFTbis}
\varphi(q)&=&C_{\rm AFT}u_1^{-1}\left({u_1\over q}\right)^2
\,I_{kl}\left(\sqrt{u_1\over q}\right) \\
&\simeq& C(k+l+2)u_1^{-1}\left[{q\over u_1}-1\right]_+^{k+l+1},
\end{eqnarray}
where we have defined the shorthands
\begin{eqnarray}
\label{IpAFT}
I_{kl}(p)&:=&\theta(1-p)\int_p^1(z-p)^k(1-z^2)^lz^{-(k+2l+5)}\,dz,
\nonumber \\ \\ 
\label{CAFT}
C_{\rm AFT}&:=&2E_0^{k+2}F_0^{l+1},
\end{eqnarray}
$u_1$ is again given by (\ref{u1E0F0}), and $C$ is given by
\begin{equation}
C=C_{\rm AFT}2^{k+1}{\Gamma(k+1)\Gamma(l+1)\over\Gamma(k+l+3)}.
\end{equation}
Note that
\begin{equation}
\left({d\over dp}\right)^{k+1}I_{kl}(p)=k! (1-p^2)^l p^{-(2l+5)}.
\end{equation}
We have normalised $C$ so that this ansatz has the same leading term
as Ansatz~1 and 2, with the combination $k+l+1$ here corresponding to
our single parameter $k$. Note that (\ref{PhiAFT}) does not explicitly
have an overall adjustable parameter, as the authors take the approach
where $h(E,F)$ is fixed and the mass is found by shooting. Moreover,
the implied parameter $C_{\rm AFT}$ has dimension.

Translating the parameters $E_0$ and $F_0$ into our notation $u_1$ and
$C$ clarifies their role: solutions exist with $27< u_1/M^2\le 729/16$
and $k+l+2\le -1$, or $u_1/M^2=27$ and $k+l+2\le -1/2$, and the shape of
the solution (up to an overall scale) depends only on $u_1/M^2$ and
the powers $k$ and $l$. This means that solutions exist precisely for
$C_{\rm AFT}$ (interpreting this dimensional constant as dimensionless
by an implicit choice of units for $E_0$ and $F_0$) in an interval
that depends on $k$ and $l$, and within that interval the value of
$C_{\rm AFT}$ determines $u_1/M^2$.


\end{document}